\documentclass[aps,showpacs,prl,toolkits,twocolumn]{revtex4}
\usepackage{epsfig,amssymb,amsfonts,amsmath,graphicx}

\makeatletter
\renewcommand{\@makefntext}[1]{\parindent=1em\noindent\hbox to 1.8em
{\hss$^{\@thefnmark}$}#1}
\renewcommand{\@footnotemark}{\hbox{\mathsurround=0pt$^{\@thefnmark}$}}

\makeatother

\begin{document}
\title{Chiral symmetry patterns of excited mesons with the Coulomb-like linear
confinement.}
\author{R. F. Wagenbrunn    and L. Ya. Glozman}
\affiliation{Institute for
 Physics, Theoretical Physics branch, University of Graz, Universit\"atsplatz 5,
A-8010 Graz, Austria}

\newcommand{\be}{\begin{equation}}
\newcommand{\bea}{\begin{eqnarray}}
\newcommand{\ee}{\end{equation}}
\newcommand{\eea}{\end{eqnarray}}
\newcommand{\ds}{\displaystyle}
\newcommand{\low}[1]{\raisebox{-1mm}{$#1$}}
\newcommand{\loww}[1]{\raisebox{-1.5mm}{$#1$}}
\newcommand{\lmn}{\mathop{\sim}\limits_{n\gg 1}}
\newcommand{\vpint}{\int\makebox[0mm][r]{\bf --\hspace*{0.13cm}}}
\newcommand{\too}{\mathop{\to}\limits_{N_C\to\infty}}
\newcommand{\vp}{\varphi}
\newcommand{\vx}{{\vec x}}
\newcommand{\vy}{{\vec y}}
\newcommand{\vz}{{\vec z}}
\newcommand{\vk}{{\vec k}}
\newcommand{\vq}{{\vec q}}
\newcommand{\vpp}{{\vec p}}
\newcommand{\vn}{{\vec n}}
\newcommand{\vg}{{\vec \gamma}}

\begin{abstract}
The spectrum of $\bar q q$ mesons in a model where the only interaction
is a linear Coulomb-like instantaneous confining potential is studied. 
The single-quark Green function as well as the dynamical chiral
symmetry breaking 
are obtained from the Schwinger-Dyson (gap) equation. Given the dressed 
quark propagator, a complete spectrum of "usual" mesons is obtained from the 
Bethe-Salpeter equation. The spectrum exhibits restoration of chiral and
$U(1)_A$ symmetries at large spins and/or radial excitations. This property
is demonstrated both analytically and numerically. At large spins and/or
radial excitations higher degree of degeneracy is observed,
namely all  states with the given spin fall into reducible
representations $[(0,1/2) \oplus (1/2,0)] \times [(0,1/2) \oplus (1/2,0)]$
that combine all possible chiral multiplets with the given $J$ and $n$. 
The structure of the meson
wave functions as well as the form of the angular and radial Regge trajectories
are investigated.
\end{abstract}

\maketitle

\section{Introduction}

In spite of numerous efforts for 30 years the physics which drives
the structure of hadrons in the light flavor sector is not yet
completely understood. It is clear, however, that here the most
important elements of QCD are spontaneous breaking of chiral symmetry
and confinement. The hadron spectrum is a key to understand QCD in
the nonperturbative regime, the role and interrelations of confinement
and chiral symmetry breaking.

The spontaneous breaking of chiral symmetry is crucial 
for a proper description of the
physics of the low-lying hadrons. It is obvious, for example, in the
case of a pion, which has a dual nature. From the chiral symmetry
point of view it is a (pseudo) Goldstone excitation associated with
the spontaneously broken axial part of chiral symmetry \cite{GL}. From the
microscopical point of view, it is a highly collective quark-antiquark
mode \cite{NJL}. The latter feature persists in any known microscopical
approach to chiral symmetry breaking that is consistent with the conservation
of the axial vector current.
By now it is well understood that both these views are 
complementary and consistent with each other.
That the chiral symmetry breaking is crucially important for a
proper physics of the nucleon follows e.g. from the large pion-nucleon
coupling constant or from the Ioffe formula \cite{IOFFE} that 
relates, 
though not rigorously, the nucleon mass to the quark condensate of the vacuum.

It has been realized in recent years that the physics in the upper part
of the light hadron spectra is quite different. In this case the chiral
symmetry breaking in the vacuum is almost irrelevant, as evidenced by the
persistence of the approximate multiplets of chiral and $U(1)_A$
groups both in baryon \cite{G1,CG,G2} and mesons \cite{G3} spectra,
for a short overview see \cite{G5}. This phenomenon, if confirmed
experimentally by discovery of still missing states, is referred to as
effective chiral symmetry restoration or chiral symmetry restoration
of the second kind. It should not be mixed up
with the chiral symmetry restoration at high temperatures and/or densities.  
In the latter case the system undergoes a phase transition and the chiral 
order parameter vanishes - the quark vacuum becomes trivial. In contrast,
nothing happens of course with the quark condensate of the vacuum once
we are interested in the given isolated hadron at zero temperature. If
a hadron is highly excited, this quark condensate of the vacuum becomes
simply unimportant for physics, because the valence quarks decouple
from the condensate.

A fundamental origin of this phenomenon is that in highly excited hadrons the
semiclassical regime is manifest and semiclassically the quantum
fluctuations of the quark fields are suppressed relative to the classical
contributions that preserve both chiral and $U(1)_A$ symmetries \cite{G5}.
This general claim has been illustrated \cite{G6} within the exactly solvable
chirally symmetric confining model \cite{Orsay}, which can be
considered as a generalization of the large $N_c$ 't Hooft model
in 1+1 dimensions \cite{HOOFT} to 3+1 dimensions.
The key issue is that the chiral symmetry breaking Lorentz-scalar
dynamical mass of quarks arises via  loop dressing and vanishes
at large momenta, where only classical contributions survive. When
one increases the excitation energy of a hadron, one also increases
the typical momentum of the valence quarks. Consequently, the chiral
symmetry violating dynamical mass of quarks becomes small and chiral
symmetry gets approximately restored in the given hadron. This theoretical
expectation has been proven by the direct calculation of the heavy-light
excited meson spectra with the quadratic confining potential \cite{KNR} and
of the light-light meson spectra with the linear potential \cite{WG}.

 The effective chiral restoration has transparently been
illustrated within a simple toy model at the hadron level
that generalizes the sigma model 
and contains
an infinite amount of excited pions and sigmas \cite{CG2}.

Restoration of chiral
symmetry requires that hadrons should decouple from the Goldstone
bosons \cite{G2,CG2,JPS,G7,SV,NRS}. A hint for such a decoupling is well
seen: The coupling constant for the decay process $h^* \rightarrow h + \pi$
decreases very fast once one increases the excitation energy of a hadron.

The effective restoration of chiral symmetry has been discussed
recently within  different approaches in refs. 
\cite{A,SHIFMAN,GOL,S,DEGRAND,COHEN}.

In this paper we present a formalism that leads
to the meson spectrum reported in our recent 
Letter \cite{WG} as well as some additional numerical results.
We also study meson wave functions and clarify origins of a very
fast restoration with increasing  $J$ and of a  slow restoration
with increasing  the radial quantum number $n$.

Our choice of a model is motivated by the following constraints.
The model must be (i) 3+1 dimensional, because we know that in 1+1
dimensions the effective chiral restoration does not occur due to
absence of the rotational motion and spin; (ii) chirally symmetric; 
(iii) field-theoretical in nature  in order to be able
to exhibit spontaneous breaking of chiral symmetry;  (iv) confining; 
(v) preserving the conservation of the axial vector current; and (vi) solvable.
Then, upon solving a spectrum within such a model one can 
judge whether the chiral symmetry patterns appear or not 
and whether the physical arguments presented
above are correct. 

\begin{figure}
\includegraphics[width=0.6\hsize]{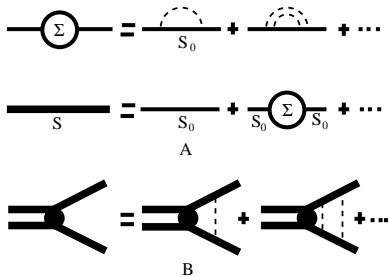}
\caption{ Graphical representation of the Schwinger-Dyson (A) and
Bethe-Salpeter equations  (B) in the ladder approximation.}
\end{figure}

The only known model that satisfies all these constraints is the
Generalized Nambu and Jona-Lasinio model of ref.   \cite{Orsay}.
In this model the only interaction between quarks is the instantaneous
confining Lorentz-vector potential between color quark charges. 
This model  belongs intrinsically  to the class of large $N_c$
models, because only ladders are taken into account upon solving
the Schwinger-Dyson and Bethe-Salpeter equations, see Fig. 1, and
there are no vacuum fermion loops.
Different aspects of this model, in particular chiral symmetry breaking, 
have been studied in many works, see for example 
\cite{Adler:1984ri,Alkofer:1988tc,BR,BN} and the spectrum of the lowest
excitations of the light-light mesons with the linear potential has
been calculated in ref. \cite{COT}. It has been recently applied to study
confinement in the diquark systems \cite{AL}.

From the underlying physics point of view, this model can be directly
related to the Gribov-Zwanziger scenario of confinement in Coulomb gauge
\cite{GRIBOV}. In Coulomb gauge a Hamiltonian setup to QCD can
be used \cite{CL,Finger:1981gm} and numerous studies in different 
approaches suggest
that the Coulomb interaction in the infrared turns out to be a
confining linear potential, see e.g. \cite{Zwanziger:2003de,Greensite:2004ke,
SS,R}. Actually the infinitely increasing Coulomb potential is a necessary
but not a sufficient condition for confinement mechanism in QCD
and the strength of the 
Coulomb potential
must  be larger or equal to that extracted from the Wilson loop
\cite{Zwanziger:2003de}. Then
there is no contradiction with the recent
observation on the lattice that in the deconfining phase above $T_c$ 
a strong Coulomb confinement still persists \cite{Nakamura:2005ux}. This
shows, however, that the actual confining scenario in QCD is richer than a
simple linear Coulomb-like potential. Hence some elements of
QCD are still missing in this picture and one needs them in order
to obtain a completely realistic picture of hadrons. 

However, our purpose is different. We use this model
to answer a principal question whether effective restoration
of chiral symmetry in excited hadrons occurs or not and if it does -
to use the model as a laboratory to get an insight.
This model satisfies all the criteria (i)-(vi) and hence is completely
adequate for this purpose.

\section{ The Hamiltonian and  the infrared regularization}

The GNJL model is described by the Hamiltonian \cite{Orsay}
\begin{eqnarray} 
\hat{H} & = & \int d^3x\bar{\psi}(\vec{x},t)\left(-i\vec{\gamma}\cdot
\vec{\bigtriangledown}+m\right)\psi(\vec{x},t) \nonumber \\
 &+& \frac12\int d^3
xd^3y\;J^a_\mu(\vec{x},t)K^{ab}_{\mu\nu}(\vec{x}-\vec{y})J^b_\nu(\vec{y},t),
\label{H} 
\end{eqnarray} 
%
with the quark current-current interaction
\begin{equation}
J_{\mu}^a(\vec{x},t)=\bar{\psi}(\vec{x},t)\gamma_\mu\frac{\lambda^a}{2}
\psi(\vec{x},t)
\end{equation}
parametrized by an instantaneous 
confining kernel $K^{ab}_{\mu\nu}(\vec{x}-\vec{y})$ of a generic
form. We use the linear confining potential,
\begin{equation} 
K^{ab}_{\mu\nu}(\vec{x}-\vec{y})=g_{\mu 0}g_{\nu 0}
\delta^{ab} V (|\vec{x}-\vec{y}|),
\label{KK}
\end{equation}
%
and absorb the color Casimir factor into the Coulomb string tension,

\begin{equation}
\frac{\lambda^a \lambda^a}{4}V(r) = \sigma r.
\label{st}
\end{equation}

\noindent
The Fourier transform as well as
the loop integrals calculated with the linear potential are
infrared divergent. Hence we have to perform the infrared regularization
and suppress the small momentum behavior of the linear potential
by introducing a cutoff parameter into the potential. Then in the
final answer this cutoff parameter must be sent to 0, i.e. the
infrared limit must be taken.
While quantities, such as the single quark Green
function, can be divergent in the infrared limit, which means
that a single quark cannot be observed,
we have to check that any color-singlet observable, like meson mass,
 is  finite and be not dependent on
the infrared regulator in the infrared limit.

There are slightly different and physically equivalent ways  to perform
the infrared regularization in the literature. We    {\it define} the
potential in momentum space as in ref \cite{Alkofer:1988tc}
\begin{equation}
V(p)= \frac{8\pi\sigma}{(p^2 + \mu_{\rm IR}^2)^2}.
\label{FV} 
\end{equation}
Then, upon the transformation back into a configurational
space 
\begin{eqnarray}
- \frac{1}{(2\pi)^3}\int d^3p \;e^{i\vec p \vec r} V(p) &=&
-\frac{\sigma\exp(-\mu_{\rm IR}r)}{\mu_{\rm IR}}\nonumber\\&=& 
 - \frac{\sigma}{\mu_{\rm IR}}+\sigma r+{\cal O}(\mu_{\rm IR})
\label{div}
\end{eqnarray}
%
one recovers that the potential contains the required linear
potential plus a term that diverges in the infrared limit 
$\mu_{\rm IR} \rightarrow 0$. Note that the divergent part is canceled
exactly in all physical observables and there is no dependence on
the particular way of the infrared regularization.

 Given this prescription all the
integrals are well defined and we can proceed by solving the
Schwinger-Dyson (gap) and Bethe-Salpeter equations. Note that there are
no ultraviolet divergences if only a linear potential is used. They will
persist, however, once a Coulomb potential is added. Then 
the ultraviolet regularization and renormalization would be required. 
Since our purpose is to study the chiral symmetry
restoration in the highly excited  light-light states, where a role
of the Coulomb interaction is negligible, such a complication is not required.

\section{The gap equation and the chiral symmetry breaking}

Due to interaction with the gluon field the dressed Dirac operator for
the quark becomes
\begin{equation}
D(p_0,\vec{p})= i S^{-1}(p_0,\vec{p}) = D_0(p_0,\vec{p})-\Sigma(p_0,\vec{p}),
\label{SAB}
\end{equation}

\noindent
where the bare Dirac operator with the bare quark mass $m$ is

\begin{equation}
D_0(p_0,\vec{p})=i S_0^{-1}(p_0,\vec{p})=p_0\gamma_0-\vec{p}\cdot\vec{\gamma}-m,
\label{bare}
\end{equation}

\noindent
 and $\Sigma(p_0,\vec{p})$
is the quark self energy. For an instantaneous model
the latter becomes energy ($p_0$) independent and is given by
\begin{equation}
i\Sigma(\vec{p})=\int\frac{d^4q}{(2\pi)^4}V(|\vec{p}-\vec{q}|)
\gamma_0\frac{1}{S_0^{-1}(q_0,\vec{q})-\Sigma(\vec{q})}\gamma_0. 
\label{Sigma03} 
\end{equation}
Inserting (\ref{Sigma03})  into (\ref{SAB}) 
yields an integral equation for $S(p_0,\vec{p})$.
With the ansatz
\begin{equation}
\Sigma(\vec{p}) =A(p)-m+\vec{\gamma}\cdot\hat{p}(B(p)-p),
\end{equation}
%
where here and in the following $p$ means $|\vec{p}|$ and $\hat{p}=\vec{p}/p$,
the equation (\ref{SAB}) can be cast into a coupled system of integral equations
\begin{eqnarray}\label{gapequdetaila}
A(p)&=&m+\frac{1}{2}\int \frac{d^3q}{(2\pi)^3}\;V(k)
\frac{M(q)}{\bar{\omega}(q)},\\\label{gapequdetailb}
B(p)&=&p+\frac{1}{2}\int \frac{d^3q}{(2\pi)^3}\;V(k)\;\hat{p}\cdot\hat{q}\;
\frac{q}{\bar{\omega}(q)}\;.
\end{eqnarray}
where the Lorentz-scalar $A(p)$ and Lorentz-spatial vector $B(p)$
parts of self-energy contain classical and quantum loop contributions \cite{G6}. 
Here
\begin{equation}
M(p)=p\frac{A(p)}{B(p)}
\end{equation}
is called the quark mass function,
$\bar{\omega}(p)=\sqrt{M^2(p)+p^2}$ and $\vec{k}=\vec{p}-\vec{q}$.
The appearance of a dynamical mass signals the spontaneous chiral symmetry breaking.
In the chiral limit $m=0$ it has exclusively
a quantum (loop) origin.

Notice that in (\ref{gapequdetaila},\ref{gapequdetailb}) the integrals
are divergent at $k=0$ unless an infrared
(IR) regulator $\mu_{\rm IR}>0$ is introduced. 
By defining the chiral angle $\varphi_p$ as
\begin{eqnarray}
\sin\varphi_p=&\displaystyle\frac{A(p)}{\omega(p)}=&\frac{M(p)}{\bar{\omega}(p)},\\
\cos\varphi_p=&\displaystyle\frac{B(p)}{\omega(p)}=&\frac{p}{\bar{\omega}(p)},
\end{eqnarray} 
where 
\begin{equation}
\omega^2(p)={A^2(p)+B^2(p)},
\label{eq:omega}
\end{equation}
(\ref{gapequdetaila},\ref{gapequdetailb}) can be transformed into
one integral equation
\begin{equation}\begin{array}{l}
p\sin\varphi_p-m\cos\varphi_p=\\ \displaystyle
\frac{1}{2}
\int \frac{d^3q}{(2\pi)^3}\;V(k)
\left(\sin\varphi_q\cos\varphi_p-
\hat{p}\cdot\hat{q}\cos\varphi_q\sin\varphi_p\right)
\end{array}
\label{eq:gapphip}
\end{equation}
for the chiral angle.
The integral in (\ref{eq:gapphip}) is convergent
at $k=0$ since the  infrared divergences arising by
integrating over the two terms of the integrand mutually cancel. 
Hence the chiral angle and the dynamical mass

\begin{equation}
M(p)=p \tan\varphi_p,
\end{equation}
 are infrared finite. Then the quark condensate is given as

\begin{equation}
\langle\bar{q}q\rangle=-\frac{N_C}{\pi^2}\int^{\infty}_0 dp\;p^2\sin\vp_p.
\label{Sigma1}
\end{equation}

\begin{figure}[t]
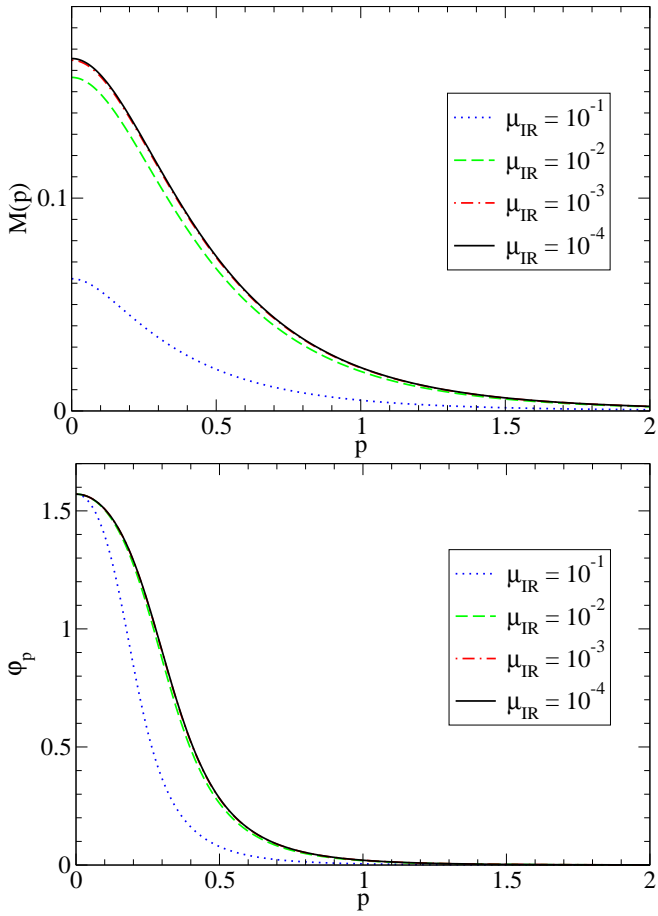

\includegraphics[width=\hsize,clip=]{m.eps}
\includegraphics[width=\hsize,clip=]{chiralangle.eps}
\caption{Mass function and chiral angle in the chiral limit
for different values of the infrared regulator $\mu_{\rm IR}$.
All quantities are given in appropriate units of $\sqrt{\sigma}$.}
\end{figure}

On the other hand,
$A(p)$, $B(p)$ and consequently also $\omega(p)$ diverge like 
$\mu_{\rm IR}^{-1}$.
The latter can be determined from $A(p)$ and $B(p)$ with Eq. (\ref{eq:omega})
or directly by combining (\ref{gapequdetaila},\ref{gapequdetailb}) to
\begin{equation}
\begin{array}{l}
\omega(p)=m\sin\varphi_p+p\cos\varphi_p+\\ \displaystyle
\frac{1}{2}
\int \frac{d^3q}{(2\pi)^3}\;V(k)
\left(\sin\varphi_q\sin\varphi_p+
\hat{p}\cdot\hat{q}\cos\varphi_q\cos\varphi_p\right).
\end{array}
\label{eq:omegaint}
\end{equation}
The infrared divergences can be explicitly projected out by using
\begin{equation}
\begin{array}{l}
\displaystyle\lim\limits_{\mu_{\rm IR}\to 0}
\frac{\mu_{\rm IR}}{\pi^2}\int d^3q\;\frac{1}{(k^2+\mu_{\rm IR}^2)^2}
f(\vec{q})=\\ \displaystyle
\int d^3q\;\delta(\vec{p}-\vec{q})f(\vec{q})=f(\vec{p}).
\end{array}
\label{eq:irdiv}
\end{equation}
This yields
\begin{eqnarray}
A(p)&=&\frac{\sigma}{2\mu_{\rm IR}}\sin\varphi_p+A_f(p),\\
B(p)&=&\frac{\sigma}{2\mu_{\rm IR}}\cos\varphi_p+B_f(p),\\
\omega(p)&=&\frac{\sigma}{2\mu_{\rm IR}}+\omega_f(p),
\label{eq:iromega}
\end{eqnarray}
where $A_f(p)$, $B_f(p)$, and $\omega_f(p)$ are infrared-finite functions.

Numerically the integration in
(\ref{eq:gapphip}) with $\mu_{\rm IR}=0$
is feasible but demanding. Alternatively one can
also solve $(\ref{gapequdetaila},\ref{gapequdetailb})$
with sufficiently small but finite values for $\mu_{\rm IR}=0$.
Still, in particular for very small values of $\mu_{\rm IR}$,
special care has to be taken for the numerical integration
in the vicinity of $\vec{q}=\vec{p}$.
The convergence of $M(p)$ for $\mu_{\rm IR}\to 0$
in the chiral limit ($m=0$) has been 
demonstrated in Refs. \cite{Alkofer:1988tc,AL}. 
The mass function and the chiral angle, which are in agreement
with the previous studies,
are shown in the upper and lower plot of Fig. 1, respectively.
The result for
$\mu_{\rm IR}=10^{-4}\sqrt{\sigma}$ is already a very good
approximation of the infrared limit which can be determined numerically
by extrapolating the results to $\mu_{\rm IR}=0$.

An important issue is a momentum dependence of the chiral angle and
dynamical mass. Their nonzero values are uniquely related
in the chiral limit with the spontaneous symmetry breaking.
The effect of symmetry breaking is large at low momenta and
the chiral angle goes to $\pi/2$ at zero momenta.
At large momenta the mass function and the chiral angle go to zero.
This property is crucial for a proper understanding of chiral
symmetry restoration in excited hadrons.

\section{Chiral and $U(1)_A$ symmetry properties of the
noninteracting two-quark amplitude}

Given a dressed quark Green function, one can obtain a spectrum
of the quark-antiquark bound states from the homogeneous 
Bethe-Salpeter equation in the rest frame.

Before solving the Bethe-Salpeter equation we would like to discuss
a limiting case when the self-interaction is switched off, $M(p) = 0$. In
such a case there is no  self-energy and the quark condensate is
identically zero. The
chiral and $U(1)_A$ symmetries are not broken. Then the Bethe-Salpeter
amplitude is reduced to a system of massless quarks
with definite chirality. In such a case properties of the system are
unambiguously determined by the Poincar\'{e} invariance as well as
chiral and $U(1)_A$ symmetries \cite{G3}. The Poincar\'{e} invariance
prescribes the standard quantum numbers $J^{PC}$. The chiral and $U(1)_A$
properties of the system then are uniquely determined by the index
of a representation of $SU(2)_L \times SU(2)_R$ that is compatible
with the given  $J^{PC}$ and isospin $I$. Then a complete list of
states is

\begin{center}
{\bf J~=~0}

\begin{eqnarray}
(1/2,1/2)_a  &  :  &   1,0^{-+} \longleftrightarrow 0,0^{++}  \nonumber \\
(1/2,1/2)_b  &  : &   1,0^{++} \longleftrightarrow 0,0^{-+} ,
\label{sym1}
\end{eqnarray}
\bigskip
{\bf J~=~2k,~~~k=1,2,...}
\begin{eqnarray}
 (0,0)  & :  &   0,J^{--} \longleftrightarrow 0,J^{++}  \nonumber \\
 (1/2,1/2)_a  & : &   1,J^{-+} \longleftrightarrow 0,J^{++}  \nonumber \\
 (1/2,1/2)_b  & : &   1,J^{++} \longleftrightarrow 0,J^{-+}  \nonumber \\
 (0,1) \oplus (1,0)  & :  &   1,J^{++} \longleftrightarrow 1,J^{--} 
\label{sym2}
\end{eqnarray}

\bigskip
{\bf J~=~2k-1,~~~k=1,2,...}

\begin{eqnarray}
 (0,0)  & :  &   0,J^{++} \longleftrightarrow 0,J^{--}  \nonumber \\
 (1/2,1/2)_a  & : &   1,J^{+-} \longleftrightarrow 0,J^{--}  \nonumber \\
 (1/2,1/2)_b  & : &   1,J^{--} \longleftrightarrow 0,J^{+-}  \nonumber \\
 (0,1) \oplus (1,0)  & :  &   1,J^{--} \longleftrightarrow 1,J^{++} 
\label{sym3}
\end{eqnarray}
\end{center}

\noindent
The sign $\longleftrightarrow$ indicates that both given states belong to
the same representation and must be degenerate. Note, that the structure
of the chiral multiplets is different for mesons with different spin. In
particular, for some of the $J^{PC}$ the chiral symmetry requires a doubling
of states.

The $U(1)_A$ multiplets combine the opposite spatial parity states
from the distinct $(1/2,1/2)_a$ and $(1/2,1/2)_b$ multiplets of
$SU(2)_L \times SU(2)_R$.

The energy of the rotating massless quark is the same irrespective
whether it is
right or left, because there is no spin-orbit force in this case \cite{G2}.
Then one expects a degeneracy of all chiral multiplets with the same $J$.
This means that the states with the same $J$ fall into a reducible
representation 

\begin{equation}
 \left [(0,1/2) \oplus (1/2,0) \right ] \times 
\left [(0,1/2) \oplus (1/2,0) \right ],
\label{qq}
\end{equation}

\noindent
that combines all possible chiral representations of the quark-antiquark
system with the same spin.

Certainly, once the self-energy is switched on, the chiral
symmetry gets broken and there cannot be exact chiral and $U(1)_A$
multiplets. The effective restoration of chiral symmetry in excited
hadrons is defined to occur if the following conditions are satisfied:
(i) the states
fall into approximate multiplets of $SU(2)_L \times SU(2)_R$  and $U(1)_A$
and the splittings within the multiplets ( $\Delta \mu = \mu_+ -\mu_-$) 
vanish at $n \rightarrow \infty$ and/or $ J \rightarrow \infty$ ;
(ii) the splitting within the multiplet is much smaller
than between the two subsequent multiplets.

To verify this prediction we have to solve the Bethe-Salpeter equation,
which will be done in the following sections.

\section{Bethe-Salpeter equation for mesons and cancellation of the
infrared divergences}

The homogeneous Bethe--Salpeter equation (BSE)
for a quark-antiquark bound
state with total and relative four momenta $P$ and $p$, respectively, can
generally be written in covariant form as
\begin{eqnarray}
\chi(P,p)&= &- i\int\frac{d^4q}{(2\pi)^4}K(P,p,q)\;
S(q+P/2) \nonumber \\
& \times & \chi(P,q)S(q-P/2),
\label{GenericSal1}
\end{eqnarray}
where $K(P,p,q)$ is called the Bethe-Salpeter kernel and 
$\chi(P,p)$ is the meson vertex function.
In the rest frame, i.e. with the four momentum
$P^\nu=(\mu,\vec{P}=0)$ the latter depends
on the meson mass $\mu$ and all four components of $p$.
Finally, for the instantaneous interaction of our model
it is energy independent and BSE becomes
\begin{eqnarray}
\chi(\mu,\vpp)&= &- i\int\frac{d^4q}{(2\pi)^4}V(|\vpp-\vq|)\;
\gamma_0 S(q_0+\mu/2,\vk) \nonumber \\
& \times & \chi(\mu,\vq)S(q_0-\mu/2,\vk)\gamma_0.
\label{GenericSal2}
\end{eqnarray}

In order to solve the Bethe-Salpeter equation for mesons with 
the set of quantum numbers $J^{PC}$ with the
given total spin $J$,  parity $P$, and C-parity $C$, 
we expand the meson vertex function  $\chi^{PC}_{JM}(\mu,\vpp)$
into a set of all possible independent Poincar\'{e}-invariant
amplitudes consistent with $J^{PC}$. This is done in the Appendix A.
Then the Bethe-Salpeter equation transforms into a system of coupled
integral equations. Such a system of equations is derived in the Appendix B.
The expansion of the meson vertex functions as well as the system of coupled
equations look slightly differently for three possible categories of
$\bar q q$ mesons.

Category 1. To this category mesons with $J^{-+}$ for $J=2n$
and $J^{+-}$ for $J=2n+1$ belong. In the nonrelativistic limit these mesons
have wave functions with orbital angular momentum
$L=J$ and the sum of the two quark spins $S=0$.

Category 2. Mesons with $J^{++}$ for $J=2n$
and $J^{--}$ for $J=2n+1$ are in this category. In this case the nonrelativistic
wave functions have $L=J\pm 1$ and $S=1$.

Category 3. This category contains mesons with $J^{--}$ for $J=2(n+1)$
and $J^{++}$ for $J=2n+1$. Their nonrelativistic wave functions
have $L=J$ and $S=1$.

The relativistic quark-antiquark states with $J^{PC}=0^{--}$
do not exist in the BSE framework with an instantaneous interaction.
In the nonrelativistic limit
$L=0$ and $S=1$ would have to couple to $J=0$, which is impossible.
The quark-antiquark states of category 4, i.e., with $J^{+-}$
for $J=2n$ and $J^{-+}$ for $J=2n+1$,
do not exist in the BSE framework with the instantaneous interaction, too, 
as is also shown in Appendix B. 
There are also no corresponding nonrelativistic
states. Mesons with such quantum numbers are called 
exotic (mesons with exotic quantum numbers) and
can be realized in form of 
hybrids, i.e., as $q\bar q$ states with explicit
excitation of the gluonic field,
or as four-quark states.
If they can also be described as quark-antiquark states in the
the BSE framework with a non-instantaneous interaction
is a matter of dicsussion.
We do not consider them in the present paper, however.

Now we want to study the infrared limit of the Bethe-Salpeter
equation and cancellation of the infrared singularities.
To this end let us
take a closer look at the integral equations for mesons of category 1,
which are 
\begin{widetext}
\begin{subequations}
\label{eq:inteqtype1}
\begin{eqnarray}
\omega(p)h(p)&=&\frac{1}{2}\int \frac{d^3q}{(2\pi)^3} V(k)
P_J(\hat{p}\cdot\hat{q})
\left(h(q)+\frac{\mu^2}{4\omega(q)}g(q)\right),
%
\label{eq:inteqtype1a}
\\
\left(\omega(p)-\frac{\mu^2}{4\omega(p)}\right)g(p)&=&h(p)
\nonumber \\&+&\frac{1}{2}\int \frac{d^3q}{(2\pi)^3} V(k)
\frac{M(p)M(q)P_J(\hat{p}\cdot\hat{q})+pq\left(
\frac{J+1}{2J+1}P_{J+1}(\hat{p}\cdot\hat{q})
+\frac{J}{2J+1}P_{J-1}(\hat{p}\cdot\hat{q})\right)}{\bar{\omega}(p)
\bar{\omega}(q)}
g(q).
\label{eq:inteqtype1b}
\end{eqnarray}
\end{subequations}
\end{widetext}
Here the functions $h(p),g(p)$ are
as defined
in the Appendix B which along with the meson mass $\mu$ have to be found
from the given equations and $\vec{k}=\vec{p}-\vec{q}$.

According to (\ref{eq:iromega}) and (\ref{eq:irdiv}) there are IR divergent terms
\begin{equation}
\frac{\sigma}{2\mu_{\rm IR}}h(p)
\end{equation}
and
\begin{equation}
\frac{\sigma}{2\mu_{\rm IR}} \int d^3q\delta(\vec{k}) P_J(\hat{p}\cdot\hat{q})h(q)=
\frac{\sigma}{2\mu_{\rm IR}}h(p)
\end{equation}
on the left and right hand sides of (\ref{eq:inteqtype1a}), respectively.
The term with $g(q)$ in the integrand of (\ref{eq:inteqtype1a}) comes with
a factor $1/\omega(q)={\cal O}(\mu_{\rm IR})$ and the integral 
over this term becomes the
IR finite expression
\begin{equation}
\frac{\mu^2}{4}g(p).
\end{equation}
Similarly 
there are IR divergent terms
\begin{equation}
\frac{\sigma}{2\mu_{\rm IR}}g(p)
\end{equation}
and
\begin{widetext}
\begin{equation}
\frac{\sigma}{2\mu_{\rm IR}} \int d^3q\delta(\vec{k}) 
\frac{M(p)M(q)P_J(\hat{p}\cdot\hat{q})+pq\left(
\frac{J+1}{2J+1}P_{J+1}(\hat{p}\cdot\hat{q})
+\frac{J}{2J+1}P_{J-1}(\hat{p}\cdot\hat{q})\right)}{\bar{\omega}(p)\bar{\omega}(q)}
g(q)=\frac{\sigma}{2\mu_{\rm IR}}g(p)
\end{equation}
on the left and right hand sides of (\ref{eq:inteqtype1b}), respectively.
The second term in the bracket on the left hand side has a factor
$\omega(p)^{-1}={\cal{O}}(\mu_{\rm IR})$ which vanishes in the IR limit.
On the right hand side there is an IR finite term $h(p)$.
One sees that the IR divergent terms on both sides of both equations cancel
and we end up with the coupled system of integral equations
\begin{subequations}
\label{eq:finteqtype1}
\begin{eqnarray}
\omega_f(p)h(p)&=&\frac{\mu^2}{4}g(p)+\frac{1}{2}\int \frac{d^3q}{(2\pi)^3} 
V_f(k)
P_J(\hat{p}\cdot\hat{q})h(q),
%
\label{eq:finteqtype1a}
\\
\omega_f(p)g(p)&=&h(p)
\nonumber \\&
+
&
\frac{1}{2}\int \frac{d^3q}{(2\pi)^3} V_f(k)
\frac{M(p)M(q)P_J(\hat{p}\cdot\hat{q})+pq\left(
\frac{J+1}{2J+1}P_{J+1}(\hat{p}\cdot\hat{q})
+\frac{J}{2J+1}P_{J-1}(\hat{p}\cdot\hat{q})\right)}
{\bar{\omega}(p)\bar{\omega}(q)}
g(q),
\label{eq:finteqtype1b}
\end{eqnarray}
\end{subequations}
with
\begin{equation}
V_f(k)=
V(k)-\frac{\sigma}{\mu_{\rm IR}}(2\pi)^3\delta(\vec{k}),
\end{equation}
which has a finite IR limit.

In an analogous manner all IR divergences from the integral equations for
the mesons of category 2 and 3 can be removed. The IR finite integral equations
for mesons in category 2 are
\begin{subequations}
\label{eq:finteqtype2}
\begin{eqnarray}
\displaystyle
\omega_f(p)h_1(p)&=&\displaystyle\frac{\mu^2}{4}g_1(p)+
\frac{1}{2}\int \frac{d^3q}{(2\pi)^3} V_f(k)
\left\{\left(\frac{J}{2J+1}P_{J+1}(\hat{p}\cdot\hat{q})
+\frac{J+1}{2J+1}P_{J-1}(\hat{p}\cdot\hat{q})\right)
h_1(q)\right.\nonumber \\&& \displaystyle+
\left.\frac{M(q)}{\bar{\omega}(q)}\frac{\sqrt{J(J+1)}}{2J+1}
\left(P_{J+1}(\hat{p}\cdot\hat{q})-P_{J-1}(\hat{p}\cdot\hat{q})\right)
h_2(q)\right\},
\label{eq:finteqtype2a}
\end{eqnarray}
\begin{eqnarray}
%
\displaystyle
\omega_f(p)g_1(p)&=&h_1(p)
+\displaystyle\frac{1}{2}\int \frac{d^3q}{(2\pi)^3} V_f(k)\left\{
\frac{pqP_J(\hat{p}\cdot\hat{q})+M(p)M(q)\left(
\frac{J}{2J+1}P_{J+1}(\hat{p}\cdot\hat{q})
+\frac{J+1}{2J+1}P_{J-1}(\hat{p}\cdot\hat{q})\right)}
{\bar{\omega}(p)\bar{\omega}(q)}
g_1(q)\right.\nonumber 
\\&& \displaystyle+
\left.\frac{M(p)}{\bar{\omega}(p)}\frac{\sqrt{J(J+1)}}{2J+1}
\left(P_{J+1}(\hat{p}\cdot\hat{q})-P_{J-1}
(\hat{p}\cdot\hat{q})\right)g_2(q)\right\}
\label{eq:finteqtype2b}
\end{eqnarray}
\begin{eqnarray}
%
\displaystyle
\omega_f(p)h_2(p)&=&\displaystyle\frac{\mu^2}{4}g_2(p)+
\frac{1}{2}\int \frac{d^3q}{(2\pi)^3} V_f(k)
\left\{\frac{pqP_J(\hat{p}\cdot\hat{q})+M(p)M(q)\left(
\frac{J+1}{2J+1}P_{J+1}(\hat{p}\cdot\hat{q})
+\frac{J}{2J+1}P_{J-1}(\hat{p}\cdot\hat{q})\right)}
{\bar{\omega}(p)\bar{\omega}(q)}\right.
h_2(q)\nonumber \\&& \displaystyle+
\left.\frac{M(p)}{\bar{\omega}(p)}\frac{\sqrt{J(J+1)}}{2J+1}
\left(P_{J+1}(\hat{p}\cdot\hat{q})-P_{J-1}(\hat{p}\cdot\hat{q})\right)
h_1(q)\right\},
\label{eq:finteqtype2c}
%
\end{eqnarray}
\begin{eqnarray}
\displaystyle
\omega_f(p)g_2(p)&=&h_2(p)
+\displaystyle\frac{1}{2}\int \frac{d^3q}{(2\pi)^3} V_f(k)\left\{\left(
\frac{J+1}{2J+1}P_{J+1}(\hat{p}\cdot\hat{q})+
\frac{J}{2J+1}P_{J-1}(\hat{p}\cdot\hat{q})\right)
g_2(q)\right.\nonumber 
\\ &&\displaystyle+
\left.\frac{M(q)}{\bar{\omega}(q)}\frac{\sqrt{J(J+1)}}{2J+1}
\left(P_{J+1}(\hat{p}\cdot\hat{q})-
P_{J-1}(\hat{p}\cdot\hat{q})\right)g_1(q)\right\}.
\label{eq:finteqtype2d}
\end{eqnarray}
\end{subequations}

Note that the number of the auxiliary functions $h_1,g_1,h_2,g_2$ as
well as the number of coupled equations in the present case is twice
larger than for mesons of category 1. This is related to the
number of independent components in $\chi^{PC}_{JM}(\mu, \vec p)$ for $J > 0$.
However, for the states $0^{++}$ within the same category only the last
two equations (\ref{eq:finteqtype2c}) - (\ref{eq:finteqtype2d}) apply,
because in this case there are only two independent functions 
$h_2,g_2$.  

For mesons in category 3 we have:

\begin{subequations}
\label{eq:finteqtype3}
\begin{eqnarray}
\omega_f(p)h(p)&=&\frac{\mu^2}{4}g(p)\nonumber \\ 
&+&\displaystyle\frac{1}{2}\int \frac{d^3q}{(2\pi)^3} V_f(k)
\frac{M(p)M(q)P_J(\hat{p}\cdot\hat{q})+
pq\left(\frac{J}{2J+1}P_{J+1}(\hat{p}\cdot\hat{q})
+\frac{J+1}{2J+1}P_{J-1}(\hat{p}\cdot\hat{q})\right)}
{\bar{\omega}(p)\bar{\omega}(q)}h(q),
\label{eq:finteqtype3a}
%
\\
\omega_f(p)g(p)&=&h(p)
+\frac{1}{2}\int \frac{d^3q}{(2\pi)^3} V_f(k)
P_J(\hat{p}\cdot\hat{q})
g(q).
\label{eq:finteqtype3b}
\end{eqnarray}
\end{subequations}
\end{widetext}

From the integral equations the normalization of
the vertex functions cannot be determined. One can derive a normalization
condition by demanding that the charge of the isovector mesons with
isospin projection $+1$ must be one, or equivalently by normalizing the
wave function by
\begin{eqnarray}
\rm{tr}\int\frac{d^3p}{(2\pi)^3}&&\!\!\!\!\!\!\left(
\psi_{+JM}^{PC}{}^\dag(\mu,\vec{p})\psi_{+JM}^{PC}(\mu,\vec{p})\right.\nonumber\\
&&\!\!\!\!\!\!\!\!\!-\left.\psi_{-JM}^{PC}{}^\dag(\mu,\vec{p})\psi_{-JM}^{PC}(\mu,\vec{p})
\right)=2\mu
\end{eqnarray}
(see the Appendix C for the definition of the wave function),
where traces in Dirac space and color space have to be taken, the latter
giving just a factor 3.
Then for mesons of categories 1 and 3 this yields
\begin{equation}
\frac{3}{4\pi}\int\frac{d^3p}{(2\pi)^3}h(p)g(p)=1.
\label{eq:norm13}
\end{equation}
For mesons of category 2 the normalization condition becomes
\begin{equation}
\frac{3}{4\pi}\sum\limits_{i=1}^2\int\frac{d^3p}{(2\pi)^3}h_i(p)g_i(p)=
{\cal N}_1+{\cal N}_2=1,
\label{normaliz}
\end{equation}
Here ${\cal N}_1$ and ${\cal N}_2$ represent the normalization
factors of two coupled amplitudes, $(h_1,g_1)$ and $(h_2,g_2)$. Since
these amplitudes are mixed by the dynamical mass of quarks, $M(p)$,
a mixing angle $\alpha_M$ can be defined
by ${\cal N}_1=\sin^2\alpha_M$ and ${\cal N}_2=\cos^2\alpha_M$. Obviously
in the limit $M(p) \rightarrow 0$ the mixing vanishes.

\section{Effective restoration of chiral symmetry for mesons}

In this section we demonstrate that in the limit of the vanishing
dynamical mass $M(p) \rightarrow 0$ the Bethe-Salpeter equation
of the previous section recovers all the anticipated chiral multiplets
(\ref{sym1})-(\ref{sym3}).

For $M(p)=0$ the integral equations (\ref{eq:finteqtype1}),
which describe mesons of category 1, i.e. , with
$J^{-+}$ for $J=2n$ and $J^{+-}$ for $J=2n+1$,
become
\begin{widetext}
\begin{subequations}
\label{eq:bseresttype1}
\begin{equation}
\omega_f(p)h(p)=\frac{\mu^2}{4}g(p)+
\frac{1}{2}\int \frac{d^3q}{(2\pi)^3} V_f(k)
P_J(\hat{p}\cdot\hat{q})h(q),
\label{eq:bseresttype1a}
\end{equation}
\begin{equation}
\omega_f(p)g(p)=h(p)
+\frac{1}{2}\int \frac{d^3q}{(2\pi)^3} 
V_f(k)
\frac{(J+1)P_{J+1}(\hat{p}\cdot\hat{q})
+JP_{J-1}(\hat{p}\cdot\hat{q})}{2J+1}
g(q).
\label{eq:bseresttype1b}
\end{equation}
\end{subequations}

For mesons of category 2, i.e., with
$J^{++}$ for $J=2n$ ($J \neq 0$) and $J^{--}$ for $J=2n+1$, the
four coupled integral equations (\ref{eq:finteqtype2}) decouple
into two independent systems of equations
\begin{subequations}
\label{eq:bseresttype2}
\begin{equation}
\omega_f(p)h_1(p)=\frac{\mu^2}{4}g_1(p)+
\frac{1}{2}\int \frac{d^3q}{(2\pi)^3} V_f(k)
\frac{JP_{J+1}(\hat{p}\cdot\hat{q})+(J+1)P_{J-1}(\hat{p}\cdot\hat{q})}{2J+1}
h_1(q),
\label{eq:bseresttype2a}
\end{equation}
\begin{equation}
\omega_f(p)g_1(p)=h_1(p)
+\frac{1}{2}\int \frac{d^3q}{(2\pi)^3} V_f(k)
P_J(\hat{p}\cdot\hat{q})g_1(q)
\label{eq:bseresttype2b}
\end{equation}
and
\begin{equation}
\omega_f(p)h_2(p)=\frac{\mu^2}{4}g_2(p)+
\frac{1}{2}\int \frac{d^3q}{(2\pi)^3} V_f(k)
P_J(\hat{p}\cdot\hat{q})
h_2(q),
\label{eq:bseresttype2c}
\end{equation}
\begin{equation}
\omega_f(p)g_2(p)=h_2(p)
+\frac{1}{2}\int \frac{d^3q}{(2\pi)^3} V_f(k)
\frac{(J+1)P_{J+1}(\hat{p}\cdot\hat{q})
+JP_{J-1}(\hat{p}\cdot\hat{q})}{2J+1}
g_2(q).
\label{eq:bseresttype2d}
\end{equation}
\end{subequations}
For the $0^{++}$ states within this category only equations
(\ref{eq:bseresttype2c})-(\ref{eq:bseresttype2d}) apply.

For the mesons of category 3, i.e., with
$J^{++}$ for $J=2n+1$ and $J^{--}$ for $J=2(n+1)$, the Bethe-Salpeter
equation takes the form
\begin{subequations}
\label{eq:bseresttype3}
\begin{equation}
\omega_f(p)h(p)=\frac{\mu^2}{4}g(p)+
\frac{1}{2}\int \frac{d^3q}{(2\pi)^3} V_f(k)
\frac{
JP_{J+1}(\hat{p}\cdot\hat{q})
+(J+1)P_{J-1}(\hat{p}\cdot\hat{q})}{2J+1}
h(q),
\label{eq:bseresttype3a}
\end{equation}
\begin{equation}
\omega_f(p)g(p)=h(p)
+\frac{1}{2}\int \frac{d^3q}{(2\pi)^3} V(k)
P_J(\hat{p}\cdot\hat{q})
g(q).
\label{eq:bseresttype3b}
\end{equation}
\end{subequations}
\end{widetext}

We see that the equations (\ref{eq:bseresttype1a}) - (\ref{eq:bseresttype1b})
become identical to the equations 
(\ref{eq:bseresttype2c}) - (\ref{eq:bseresttype2d}) as well as the
equations (\ref{eq:bseresttype2a}) - (\ref{eq:bseresttype2b}) are
identical to equations (\ref{eq:bseresttype3a}) - (\ref{eq:bseresttype3b}).
Note, that there is no isospin dependence of the interaction, i.e.
each system of coupled equations describes both I=0 and I=1 states
which are exactly degenerate.
 Then we recover that the equations above
fall into chiral multiplets (\ref{sym1})-(\ref{sym3})
and both $SU(2)_L \times SU(2)_R$ and $U(1)_A$ are manifest.
Hence
if the typical momentum of quarks is large, as it is anticipated in 
the highly-excited mesons, the spectrum should be organized 
into chiral and $U(1)_A$ multiplets , because at large
momenta dynamical mass of quarks $M(p)$ vanishes.

\section{Numerical results for spectrum}

The method of solution for the BSE used in this work
is described in Appendix D.
In Table 1 we present our results for the spectrum for the two-flavor
($u$ and $d$) mesons in the chiral limit $m=0$. Due to chiral symmetry
breaking there is a mixing between $(h_1,g_1)$ and $(h_2,g_2)$
amplitudes for mesons of category 2 (but not for the $0^{++}$ states). 
In the language of chiral representations this mixing
means that for mesons of category 2 with $J=2n +1$ the representations
$(0,0)$ and $(1/2,1/2)_a$ are mixed in the mesons $0,J^{--}$ as well as
the representations $(0,1) \oplus (1,0)$ and $(1/2,1/2)_b$ are mixed
in the mesons $1,J^{--}$. For mesons of category 2 with $J=2n, J > 0$
the representations  $(0,0)$ and $(1/2,1/2)_a$ are mixed in the
$0,J^{++}$ states and 
the representations $(0,1) \oplus (1,0)$ and $(1/2,1/2)_b$ are mixed
in the mesons $1,J^{++}$.
The mesons of the category 2 in the Table are assigned to a definite
chiral representation according to the chiral Bethe-Salpeter amplitude
which dominates in the meson wave function, which is determined by
the relative magnitude of the normalization factors 
 ${\cal N}_1$ and ${\cal N}_2$ in (\ref{normaliz}).

Note that within the
 present model there are no vacuum fermion loops. Then since
 the interaction between quarks is isospin-independent the states
 with the same $J^{PC}$ but different isospins from the distinct
 multiplets $(1/2,1/2)_a $ and $(1/2,1/2)_b$ as well as the states
 with the same $J^{PC}$ but different isospins from $(0,0)$ and
 $(0,1) \oplus (1,0)$ representations are exactly degenerate. Hence it
 is enough to show a complete set of the isovector (or isoscalar)
 states.
 
The presented results are accurate within the quoted digits at
least for states with small $J$ and for states with higher $J$
but small $n$. At larger $J$ for larger $n$ numerical errors
accumulate in the second digit after comma.

The axial anomaly is absent within the present model, because
there are no vacuum fermion loops here. Even so there are no
exact $U(1)_A$ multiplets, because this symmetry is broken also
by the quark condensate of the vacuum. The mechanism of the $U(1)_A$
breaking and restoration is exactly the same as for $SU(2)_L \times SU(2)_R$.

\begin{figure}
\begin{center}
\includegraphics*[height=5cm]{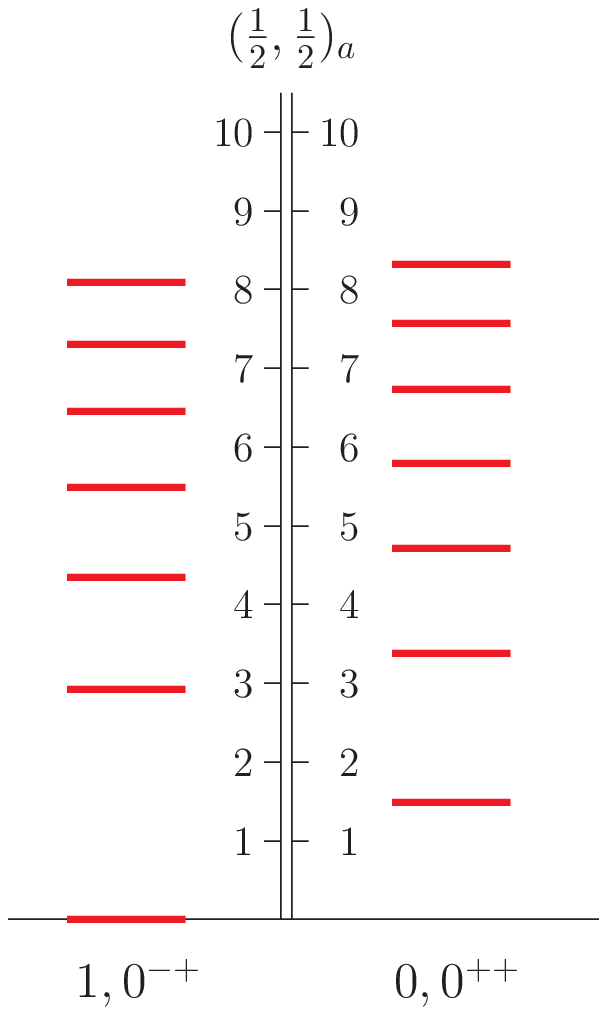}
\hspace*{1cm}
\includegraphics[height=5cm]{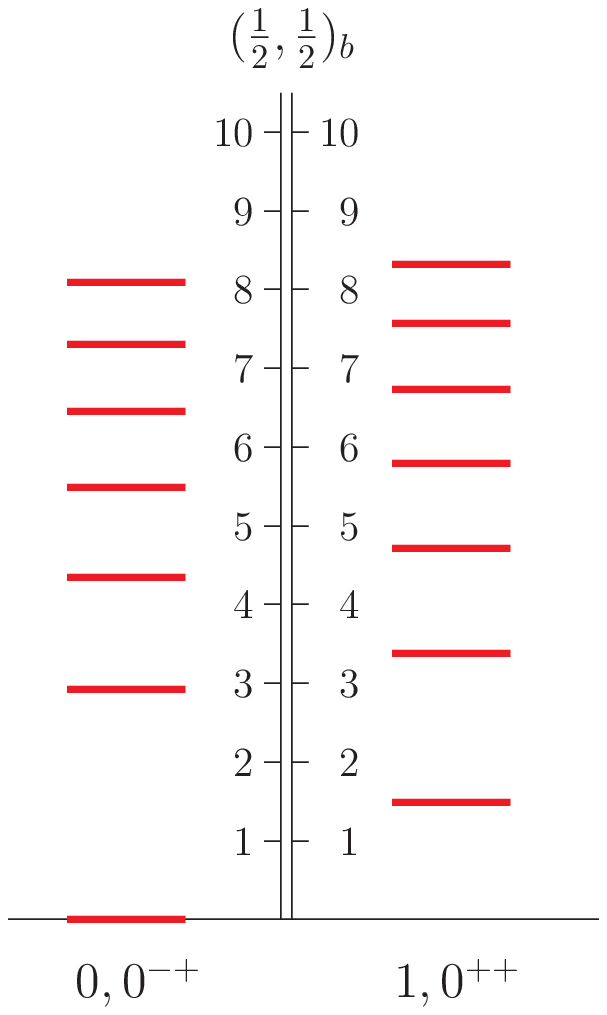}
\end{center}
\caption{ Spectra of $J=0$ mesons (masses in units of $\sqrt{\sigma}$).}
\label{s0}
\end{figure}

\begin{figure}
\begin{center}
\includegraphics[height=5cm,clip=]{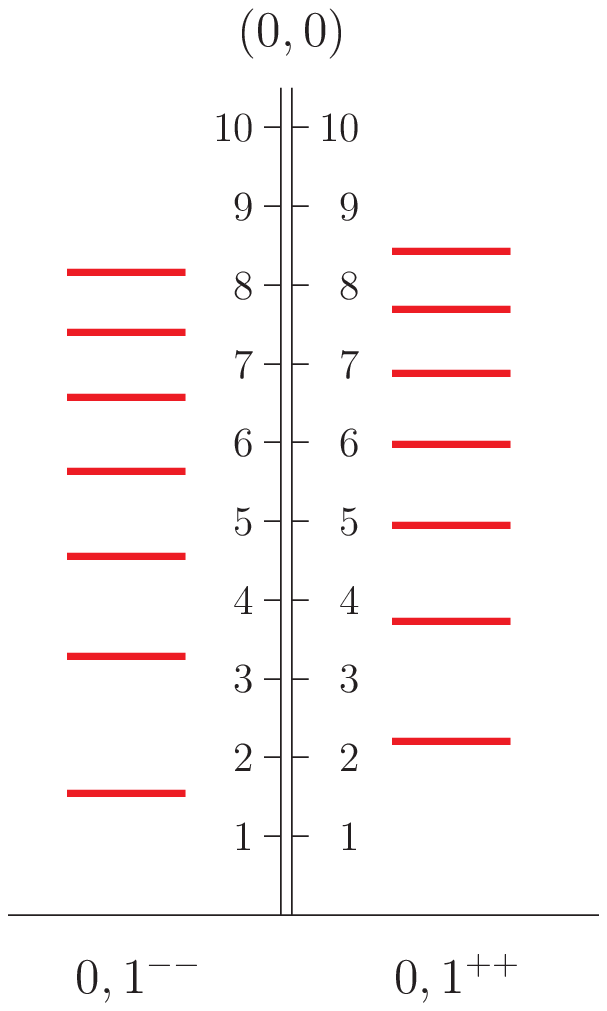}
\hspace*{1cm}
\includegraphics[height=5cm,clip=]{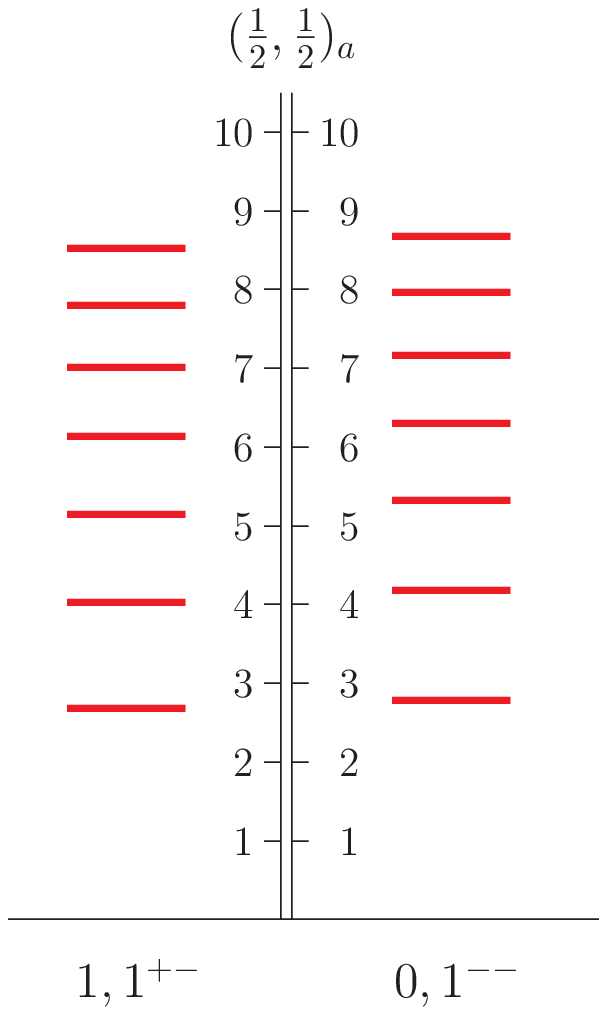}
\includegraphics[height=5cm,clip=]{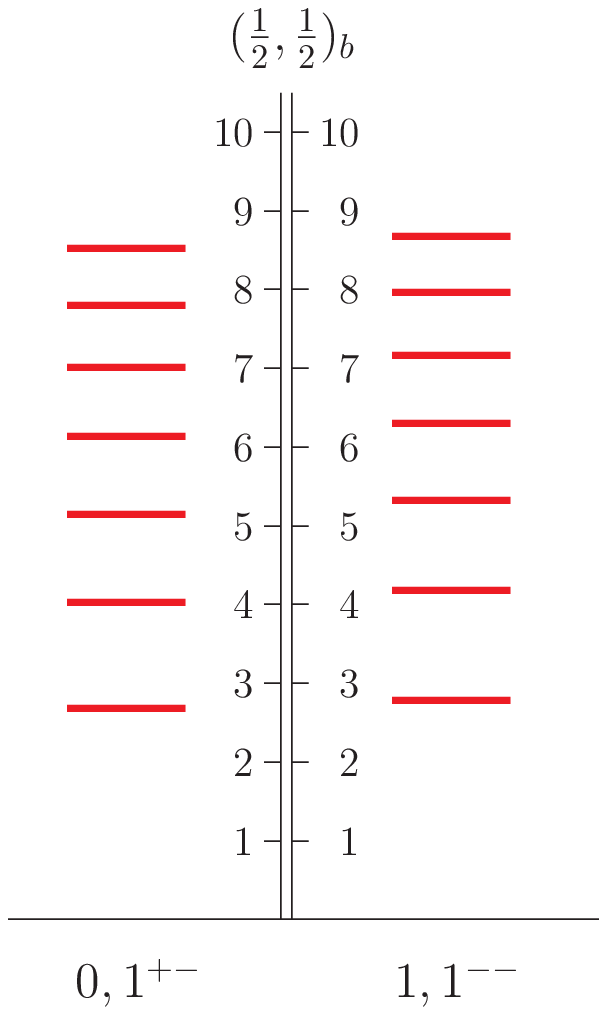}
\hspace*{1cm}
\includegraphics[height=5cm,clip=]{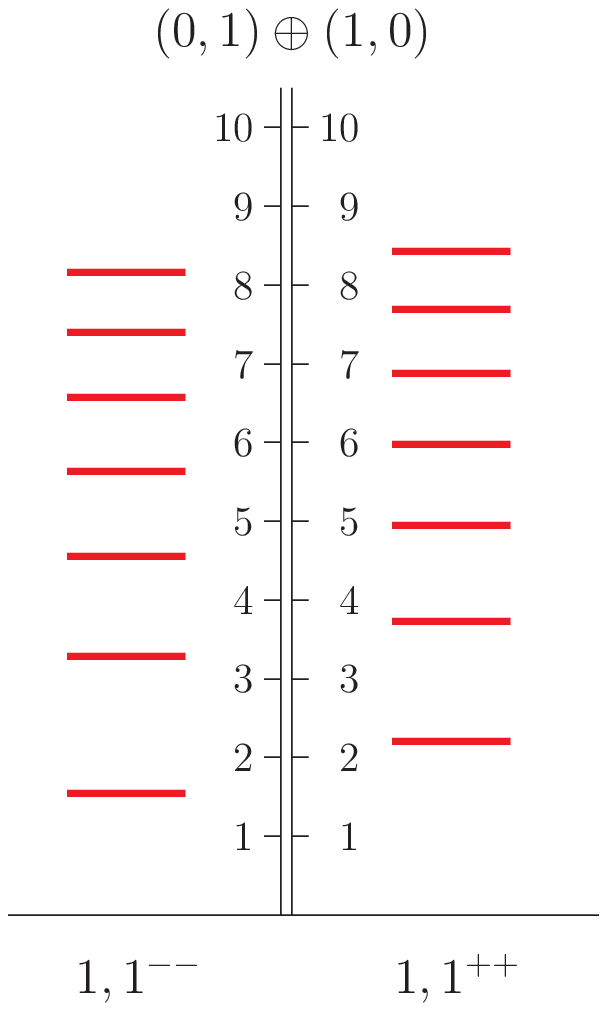}
\end{center}
\caption{ Spectra of $J=1$ mesons (masses in units of $\sqrt{\sigma}$).}
\label{s1}
\end{figure}

\begin{figure}
\begin{center}
\includegraphics[height=5cm,clip=]{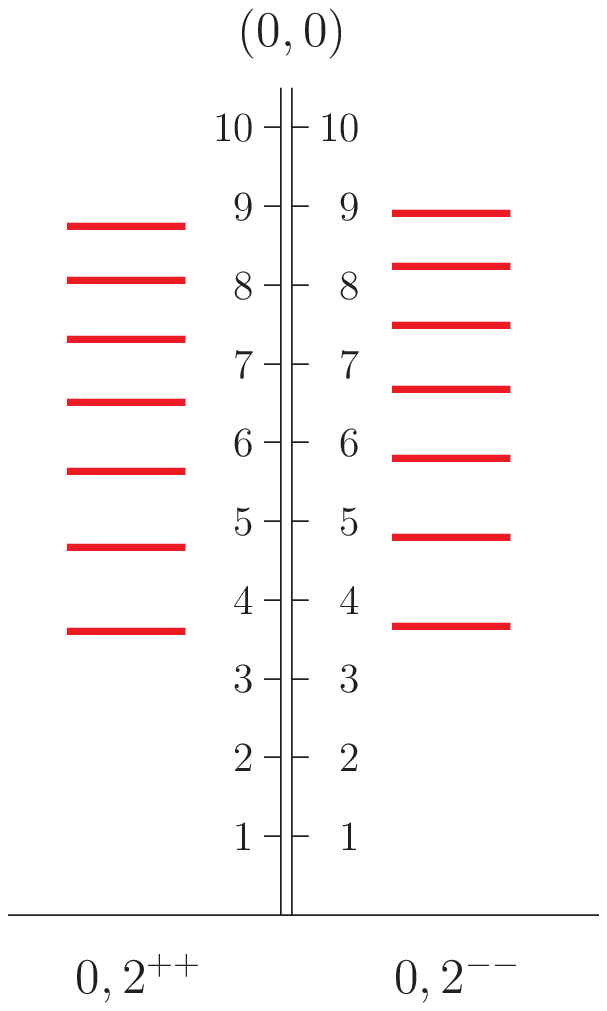}
\hspace*{1cm}
\includegraphics[height=5cm,clip=]{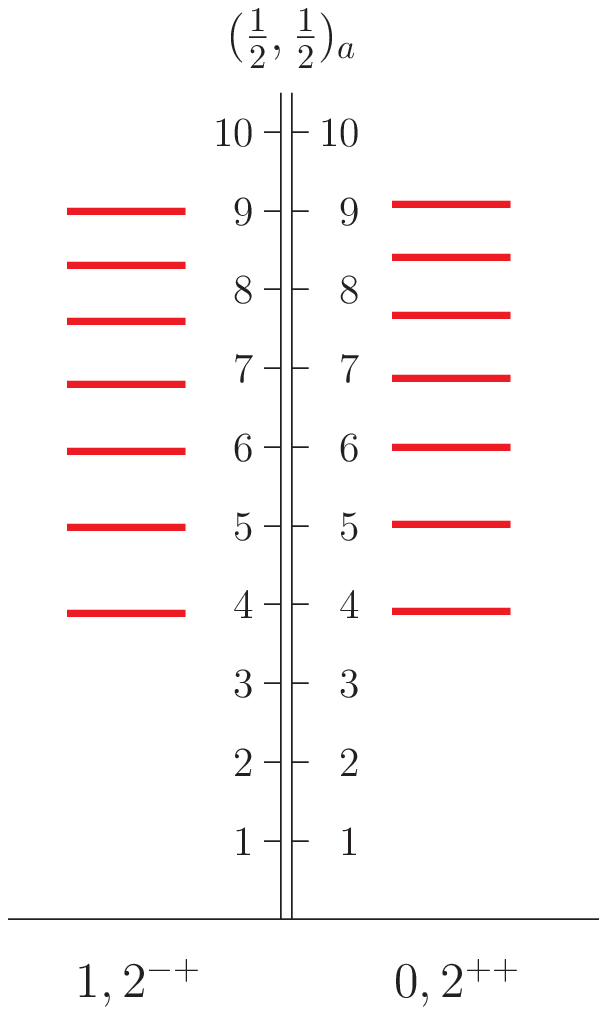}
\includegraphics[height=5cm,clip=]{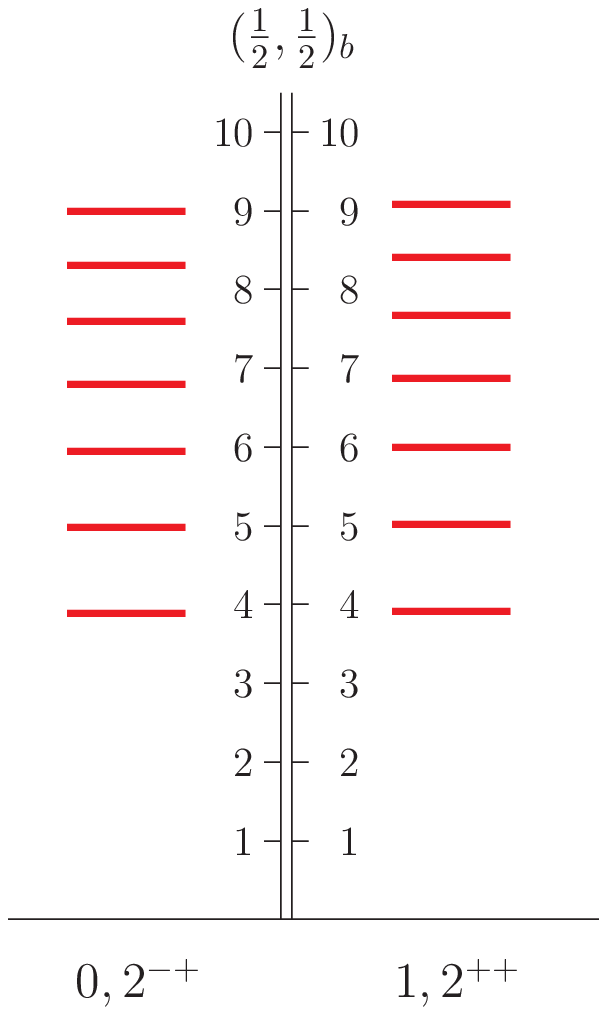}
\hspace*{1cm}
\includegraphics[height=5cm,clip=]{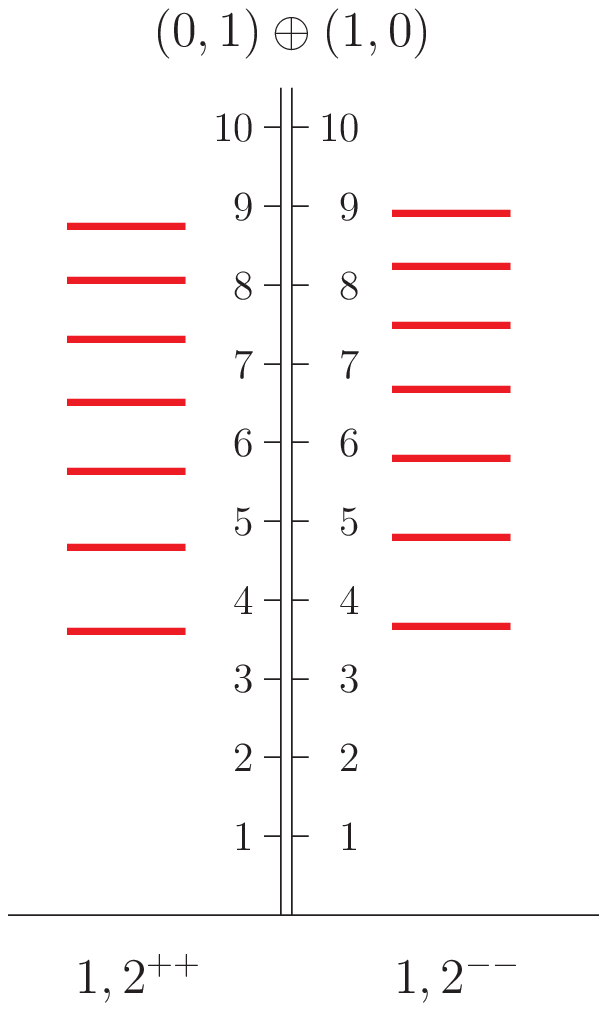}
\end{center}
\caption{ Spectra of $J=2$ mesons (masses in units of $\sqrt{\sigma}$).}
\label{s2}
\end{figure}

The excited mesons fall into approximate chiral and $U(1)_A$
multiplets and all conditions of the effective symmetry restorations
are satisfied.
We observe  a very fast restoration of
both $SU(2)_L \times SU(2)_R$ and $U(1)_A$ symmetries with increasing
$J$ and essentially more slow restoration with increasing of $n$.
\begin{table}
\caption{Masses of isovector mesons in units of $\sqrt{\sigma}$.}
\begin{tabular}{c@{\hspace*{1.5em}}c@{\hspace*{1.5em}}rrrrrrr}
\hline\hline
\multicolumn{1}{c}{chiral}&\raisebox{-1.5ex}{$J^{PC}$}&
\multicolumn{7}{c}{radial excitation $n$}\\[-1.ex]
\multicolumn{1}{c}{multiplet}&&0\hspace*{0.7em}&1\hspace*{0.7em}&2\hspace*{0.7em}&3\hspace*{0.7em}&
4\hspace*{0.7em}&5\hspace*{0.7em}&6\hspace*{0.7em}\\
\hline
$(1/2,1/2)_a$&$0^{-+}$&0.00&2.93&4.35&5.49&6.46&7.31&8.09\\
$(1/2,1/2)_b$&$0^{++}$&1.49&3.38&4.72&5.80&6.74&7.57&8.33\\
\hline
$(1/2,1/2)_a$&$1^{+-}$&2.68&4.03&5.16&6.14&7.01&7.80&8.53\\
$(1/2,1/2)_b$&$1^{--}$&2.78&4.18&5.32&6.30&7.17&7.96&8.68\\
$(0,1)\oplus(1,0)$&$1^{--}$&1.55&3.28&4.56&5.64&6.57&7.40&8.16\\
$(0,1)\oplus(1,0)$&$1^{++}$&2.20&3.73&4.95&5.98&6.88&7.69&8.43\\
\hline
$(1/2,1/2)_a$&$2^{-+}$&3.89&4.98&5.94&6.80&7.59&8.31&8.99\\
$(1/2,1/2)_b$&$2^{++}$&3.91&5.02&6.00&6.88&7.67&8.40&9.09\\
$(0,1)\oplus(1,0)$&$2^{++}$&3.60&4.67&5.64&6.51&7.32&8.06&8.75\\
$(0,1)\oplus(1,0)$&$2^{--}$&3.67&4.80&5.80&6.68&7.49&8.23&8.91\\
\hline
$(1/2,1/2)_a$&$3^{+-}$&4.82&5.77&6.62&7.41&8.13&8.81&9.45\\
$(1/2,1/2)_b$&$3^{--}$&4.82&5.78&6.65&7.44&8.17&8.86&9.50\\
$(0,1)\oplus(1,0)$&$3^{--}$&4.68&5.63&6.48&7.26&7.99&8.67&9.30\\
$(0,1)\oplus(1,0)$&$3^{++}$&4.69&5.66&6.53&7.32&8.06&8.74&9.39\\
\hline
$(1/2,1/2)_a$&$4^{-+}$&5.59&6.45&7.23&7.96&8.64&9.28&9.89\\
$(1/2,1/2)_b$&$4^{++}$&5.59&6.45&7.24&7.97&8.66&9.30&9.91\\
$(0,1)\oplus(1,0)$&$4^{++}$&5.51&6.36&7.15&7.88&8.56&9.19&9.80\\
$(0,1)\oplus(1,0)$&$4^{--}$&5.51&6.37&7.16&7.90&8.58&9.23&9.84\\
\hline
$(1/2,1/2)_a$&$5^{+-}$&6.27&7.05&7.78&8.47&9.11&9.72&10.3\\
$(1/2,1/2)_b$&$5^{--}$&6.27&7.06&7.79&8.47&9.12&9.73&10.3\\
$(0,1)\oplus(1,0)$&$5^{--}$&6.21&7.00&7.73&8.41&9.06&9.67&10.2\\
$(0,1)\oplus(1,0)$&$5^{++}$&6.21&7.00&7.73&8.42&9.07&9.68&10.3\\
\hline
$(1/2,1/2)_a$&$6^{-+}$&6.88&7.61&8.29&8.94&9.55&10.1&10.7\\
$(1/2,1/2)_b$&$6^{++}$&6.88&7.61&8.29&8.94&9.56&10.1&10.7\\
$(0,1)\oplus(1,0)$&$6^{++}$&6.83&7.57&8.25&8.90&9.51&10.1&10.7\\
$(0,1)\oplus(1,0)$&$6^{--}$&6.83&7.57&8.26&8.90&9.52&10.1&10.7\\
\hline\hline
\end{tabular}
\end{table}
\noindent
The spectrum of the $J=0,1,2$ mesons is shown in
Figs. \ref{s0} - \ref{s2}.

\begin{figure}
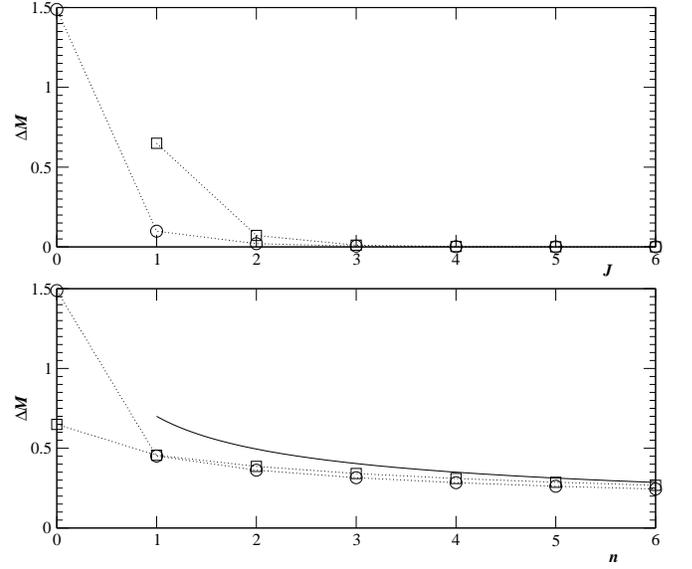

\includegraphics[width=\hsize,clip=]{delta_m_spin.eps}
\includegraphics[width=\hsize,clip=]{delta_m_radial.eps}
\caption{Mass splittings in units of $\sqrt{\sigma}$ for isovector mesons of the chiral
multiplets $(1/2,1/2)_a$ and $(1/2,1/2)_b$ (circles) and within
the multiplet $(0,1)\oplus(1,0)$ (squares) against $J$ for $n=0$ (top) and
against $n$ for $J=0$ and $J=1$, respectively (bottom).
The full line in the bottom plot is $0.7\sqrt{\sigma/n}$.}
\label{rat}
\end{figure}

In Fig. \ref{rat} the rates of the symmetry restoration against the radial
quantum number $n$ and spin $J$ are shown. It is seen that with the fixed
$J$ the splitting within the multiplets $\Delta M$ decreases asymptotically as
$1/\sqrt n$, dictated by the asymptotic linearity of the radial Regge
trajectories with different intercepts. 
With the fixed $n$ the $J$-rate of the symmetry restoration  
is much faster. 

In  the limit $n \rightarrow \infty$  or  $J \rightarrow \infty$ 
one observes a complete degeneracy of all multiplets, which means
that the states fall into 
representation  (\ref{qq}) that combines all possible chiral representations
for the systems of two massless quarks \cite{G3}. This means that in this limit the
loop effects disappear completely and the system becomes classical \cite{G5,G6}.
The analytical proof for this larger degeneracy will be given in the next
section.

\begin{figure}
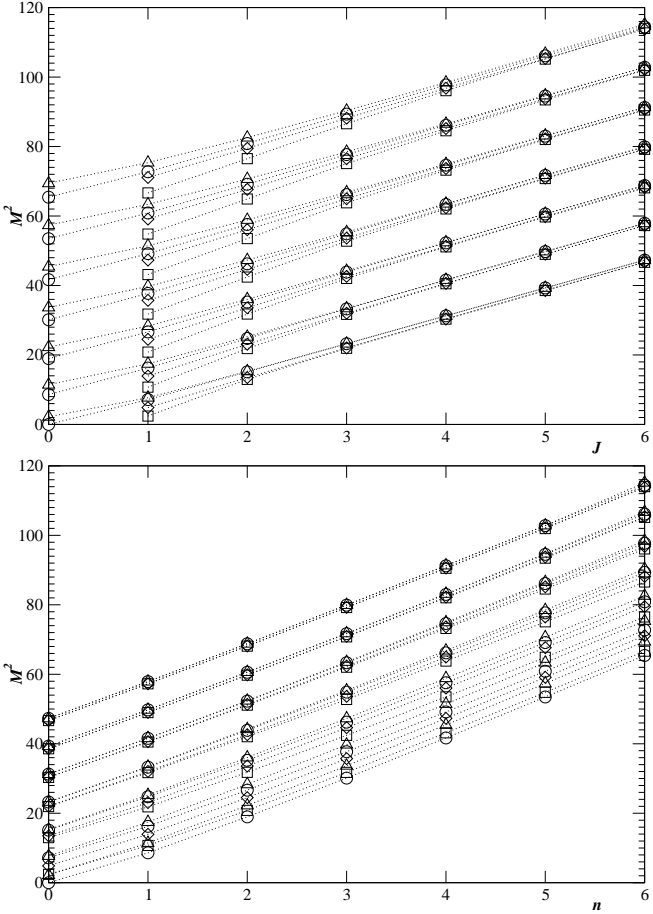

\includegraphics[width=\hsize,clip=]{regge_spin.eps}
\includegraphics[width=\hsize,clip=]{regge_radial.eps}
\caption{Angular (top) and radial (bottom) Regge trajectories for isovector mesons
with $M^2$ in units of $\sigma$.
Mesons of the chiral multiplet $(1/2,1/2)_a$ are indicated by circles,
of $(1/2,1/2)_b$ by triangles, and of $(0,1)\oplus(1,0)$
by squares ($J^{++}$ and $J^{--}$ for even and odd $J$, respectively)
and diamonds ($J^{--}$ and $J^{++}$ for even and odd $J$, respectively).
}
\label{regge}
\end{figure}

In Fig. \ref{regge} the angular and radial Regge trajectories are shown.
Both kinds of trajectories exhibit asymptotically linear behavior. This
has to be expected a-priory, because at large $J$ or $n$ all higher
Fock components are suppressed and asymptotically vanish (it will
be later well seen from the meson wave functions).Then
the semiclassical description
of the states with large $n$ or $J$ with the linear potential
requires the linear Regge trajectories, see, e.g., ref. \cite{semicl}.
There are deviations from the linear behavior at finite $J$ and $n$, however. 
This fact is obviously related to the
chiral symmetry breaking effects for lower mesons.
Note, that the chiral symmetry requires a doubling of some
of the radial and angular Regge trajectories for $J=1,2,..$. This is a
highly nontrivial prediction of chiral symmetry.
 For example, some of the
rho-mesons lie on the trajectory that is characterized by the chiral
index (0,1)+(1,0), while the other fit the trajectory with the
chiral index $(1/2,1/2)_b$. 

The  numerical result for the quark condensate is
$\langle\bar{q}q\rangle=(-0.231\sqrt{\sigma})^3$, which agrees
with the previous studies within the same model.
If we fix the string tension from the phenomenological
angular Regge trajectories, then $\sqrt{\sigma}\approx 300$ -- $400$ MeV and hence the quark
condensate is between $(-70\ \mbox{MeV})^3$ and $(-90\ \mbox{MeV})^3$ which
obviously underestimates the phenomenological value.
Probably this indicates
that other gluonic interactions could also contribute to chiral symmetry
breaking.

\section{Higher representations for excited states}

We have mentioned in the previous section that an approximate degeneracy
within the higher representation (\ref{qq}) is observed for highly excited
mesons. Below we demonstrate this property analytically.

Let us compare the integral equations (\ref{eq:finteqtype1}) 
and (\ref{eq:finteqtype3})
for mesons of categories 1 and 3, respectively. By exchanging $h$ 
and $\frac{\mu}{2}g$ for
the latter (\ref{eq:finteqtype3b}) becomes (\ref{eq:finteqtype1a}) 
and the difference between
(\ref{eq:finteqtype1b}) and (\ref{eq:finteqtype3a}) reduces to 
different coefficients 
in front of the Legendre polynomials, i.e. one has
\begin{equation}
\frac{
(J+1)P_{J+1}(\hat{p}\cdot\hat{q})
+JP_{J-1}(\hat{p}\cdot\hat{q})}{2J+1}
\label{eq1}
\end{equation} 
and
\begin{equation}
\frac{
JP_{J+1}(\hat{p}\cdot\hat{q})
+(J+1)P_{J-1}(\hat{p}\cdot\hat{q})}{2J+1}
, \label{eq2}
\end{equation} 
respectively. The same factors are also seen in eqs. 
(\ref{eq:bseresttype1a})-(\ref{eq:bseresttype3b}).
The difference between (\ref{eq1}) and (\ref{eq2}) vanishes for large $J$. 
This means
that  for large $J$ there appears degeneracy within the
whole reducible chiral multiplet
$[(0,1/2) \oplus (1/2,0)] \times [(0,1/2) \oplus (1/2,0)]$
which combines all possible chiral representations of the quark-antiquark
systems with the same $J$.

Will we see similar higher degree of degeneracy at large radial quantum
number $n$  but finite $J$? The answer is yes, which is demonstrated in 
what follows.

In general, for a large meson mass, i.e.  $n$ and/or $J$ large,
the quark-antiquark
component $\psi_+$ of the meson wave function propagating forward in time
be dominant against
the backward-propagating  component $\psi_-$ (see Appendix C). In the IR limit
$\mu/\omega(p)$ in (\ref{eq:wfpm1})--(\ref{eq:wfpm3}) goes to zero,
which means that for all three categories
of mesons $\psi_+\sim  1/2(h+\frac{\mu}{2}g)$ and 
$\psi_-\sim 1/2(h-\frac{\mu}{2}g)$.
For highly excited mesons, by neglecting the  latter and setting
\begin{equation}
\psi_+\approx  h \approx \frac{\mu}{2}g,
\end{equation}
one ends up with only one equation for $\psi_+$ for all quark-antiquark
states with the given $J$ and $n$, i.e.,
\begin{eqnarray}
\omega_f(p)\psi_+(p)&=&\frac{\mu}{2}\psi_+(p)
\nonumber\\
&+&\frac{1}{2}\int \frac{d^3q}{(2\pi)^3} V_f(k)
P_J(\hat{p}\cdot\hat{q})\psi_+(q).
\label{bselimit1}
\end{eqnarray}

\noindent
Hence all possible states with the same $J$ at the given large $n$ fall
into the $[(0,1/2) \oplus (1/2,0)] \times [(0,1/2) \oplus (1/2,0)]$ 
representation
and the degeneracy becomes exact when $J$ and/or $n$ approach infinity.

\section{Wave functions}

\begin{figure}
\includegraphics[width=\hsize,clip=]{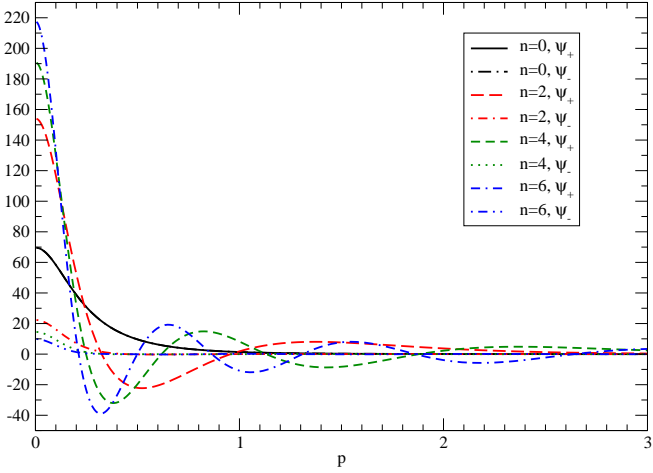}
\caption{Components $\psi_+(p)$
and $\psi_-(p)$
of the wave functions for ground and radially excited states
for mesons $0^{-+}$ of category 1.
All quantities are given in appropriate units of $\sqrt{\sigma}$.
Notice that the ground state (Goldstone boson)
 has $\mu=0$ and thus $\psi_-(p)=\psi_+(p)$, hence the two curves for $n=0$
 coincide.}
\label{fig:radialcomponents}
\end{figure}

\begin{figure}
\includegraphics[width=\hsize,clip=]{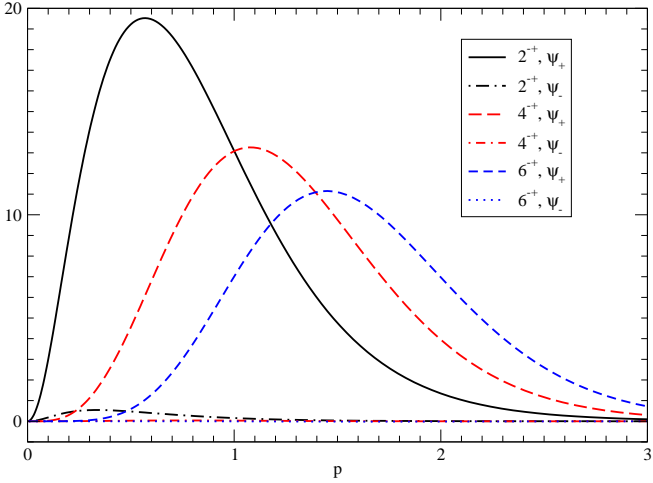}
\caption{Components $\psi_+(p)$ 
and $\psi_-(p)$
of the wave functions for mesons $J^{-+}$ 
of category 1 with radial quantum number
$n=0$ and spin $J=2,4,6$. 
All quantities are given in appropriate units of $\sqrt{\sigma}$.}
\label{fig:angularcomponents}
\end{figure}

At large $J$ or $n$ the semiclassical description requires that
the higher quark Fock components be suppressed relative the
leading one and asymptotically vanish. This is because the higher
Fock components manifestly represent effects of quantum fluctuations.
The leading Fock component is the forward-propagating quark-antiquark
component $\psi_+(p)$, defined in the Appendix C, while the higher Fock
components contain necessarily the backward-propagating quark-antiquark
states $\psi_-(p)$.
In Figs. \ref{fig:radialcomponents} and \ref{fig:angularcomponents}
we show that 
for mesons of category 1 with large radial quantum number $n$ and/or spin $J$
the forward-propagating component $\psi_+(p)$
becomes indeed much larger than the backward-propagating
component $\psi_-(p)$.
\begin{figure}
\includegraphics[width=\hsize,clip=]{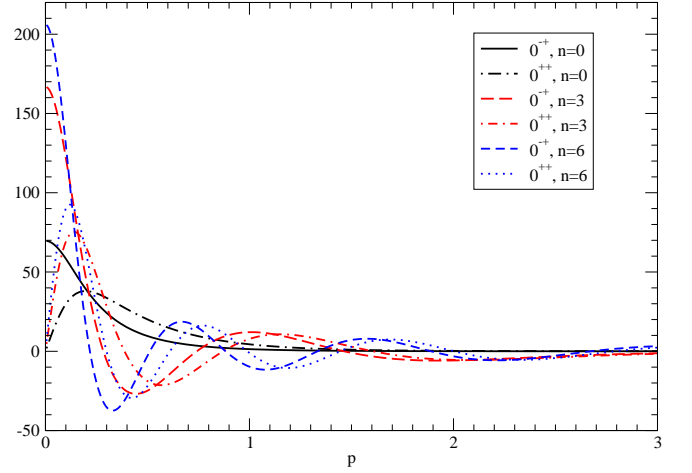}
\caption{Components $\psi_+(p)$ of
ground states and radial excitations 
of pseudoscalar and scalar mesons.
All quantities are given in appropriate units of $\sqrt{\sigma}$.}
\label{fig:ps_s}
\end{figure}
\noindent
The wave functions of mesons of categories 2 and 3 behave  similarly.
In order to study the behavior of the wave functions of mesons
with large $n$ and/or $J$ we can thus restrict ourselves
to $\psi_+(p)$.

In Fig. \ref{fig:ps_s} we compare the wave functions 
of chiral partners in $(1/2,1/2)$ representations
for mesons with 
$0^{-+}$ and $0^{++}$, 
for the ground states ($n=0$) and radial excitations $n=3,6$.
A principal difference between pseudoscalar and scalar mesons
is the behavior of the wave functions for momentum $p\to 0$ which go to a
constant for the former and to zero like $p$ for the latter.
Thus at small $p$ there is a large difference
between the wave functions of pseudoscalar and scalar mesons with the
same radial quantum number $n$ even if $n$ is large. At
large $p$ (where dynamical mass becomes small and
effective restoration of chiral symmetry takes place)
the wave functions of the radially excited pseudoscalar and scalar mesons 
behave similar but there is a shift in the positions of nodes and
maxima/minima. This persisting difference of the wave functions is a
reason for a slow rate of chiral restoration in meson 
masses with increasing $n$.

For $J>0$ there are four different mesons with the given isospin
at fixed $n$ and $J$. One of category
1 and 3, respectively, and two of category 2. 

Consider mesons of category 2.
They are
mixings of two different chiral representations with radial
functions $\psi_{1+}(p)$ and $\psi_{2+}(p)$, respectively. 
 The mixing is characterized
by the size
of  the normalization factors
${\cal N}_1$ and ${\cal N}_2=1-{\cal N}_1$.
\begin{figure}
\includegraphics[width=\hsize,clip=]{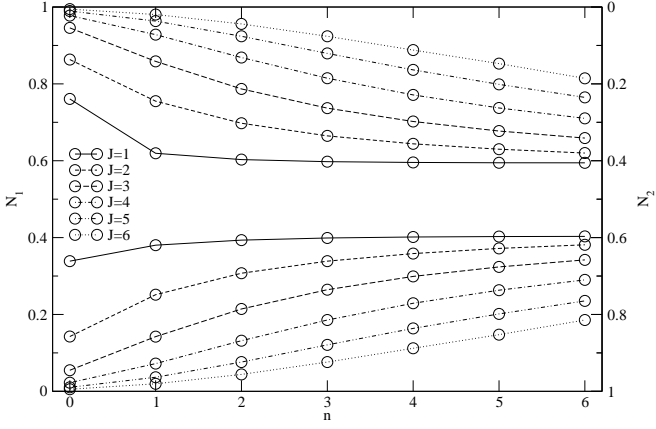}
\caption{${\cal N}_1$ (or ${\cal N}_2 = 1 - {\cal N}_1)$
for mesons of category 2. For each $J>0$ and $n$
there is one state with ${\cal N}_1>{\cal N}_2=1-{\cal N}_1$ and vice versa.}
\label{fig:mixing}
\end{figure}
In Fig. \ref{fig:mixing} we show  ${\cal N}_1$  $({\cal N}_2)$
for all states  of category 2 considered in this work.
For given $n$ the mixing decreases with increasing $J$.
At large $J$ there is one 
almost unmixed state
with ${\cal N}_1\approx 1$ and one almost unmixed state with 
${\cal N}_2\approx 1$.
The former belongs to the chiral multiplet $(0,0)$ and $(0,1)\oplus(1,0)$ 
in case
of isospin $0$ and $1$, respectively and the latter to
$(\frac{1}{2},\frac{1}{2})_a$ and $(\frac{1}{2},\frac{1}{2})_b$ in case
of isospin $0$ and $1$, respectively.
In contrast,
for given $J$ the mixing first increases with increasing $n$
and then saturates. The saturation value
seems to be of about 0.6 for $J=1$. For larger $J$ such
a saturation is not yet reached for $n=6$.
\begin{figure}
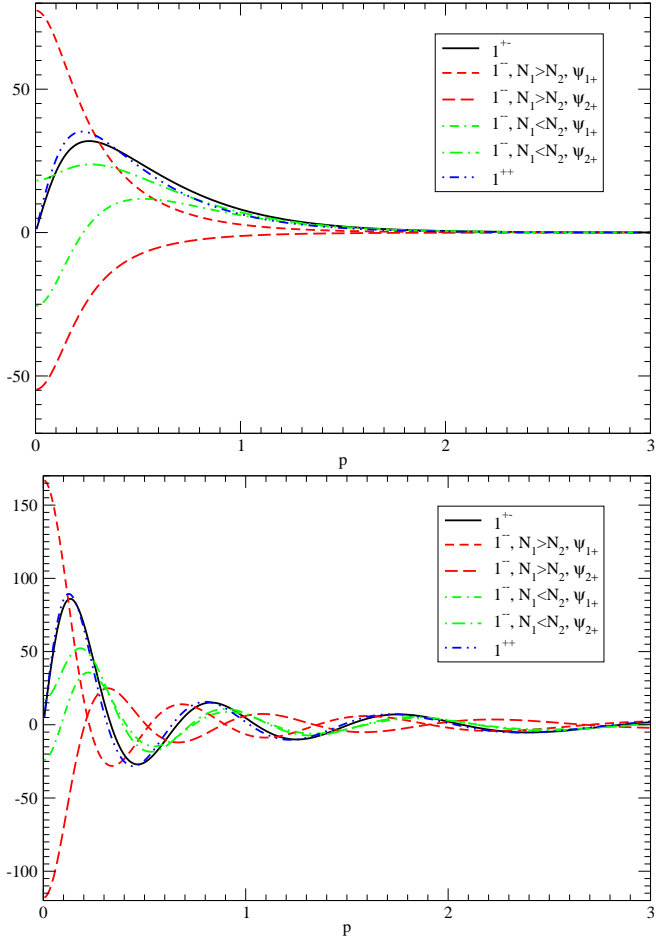

\includegraphics[width=\hsize,clip=]{J_1_wf_+.eps}
%
%
\includegraphics[width=\hsize,clip=]{J_1_6_wf_+.eps}
\caption{Components $\psi_+(p)$ of mesons with $J=1$. Upper plot: $n=0$,
lower plot: $n=6$.
All quantities are given in appropriate units of $\sqrt{\sigma}$.}
\label{fig:J1}
\end{figure}

In Fig. \ref{fig:J1} we show the wave functions 
For $J=1$ and $n=0$ and $n=6$. In both cases the wave functions of 
mesons of category 1 and 3 
are rather similar to each other. These wave functions go to zero
for $p\to 0$ contrary to the wave functions of the mesons of category 2 
which go
to a constant. At small momenta the coupling between  
$\psi_{1}(p)$ and $\psi_{2}(p)$ in mesons of category 2
becomes large and thus there is a rather strong mixing even for high radial
excitations.

\begin{figure}
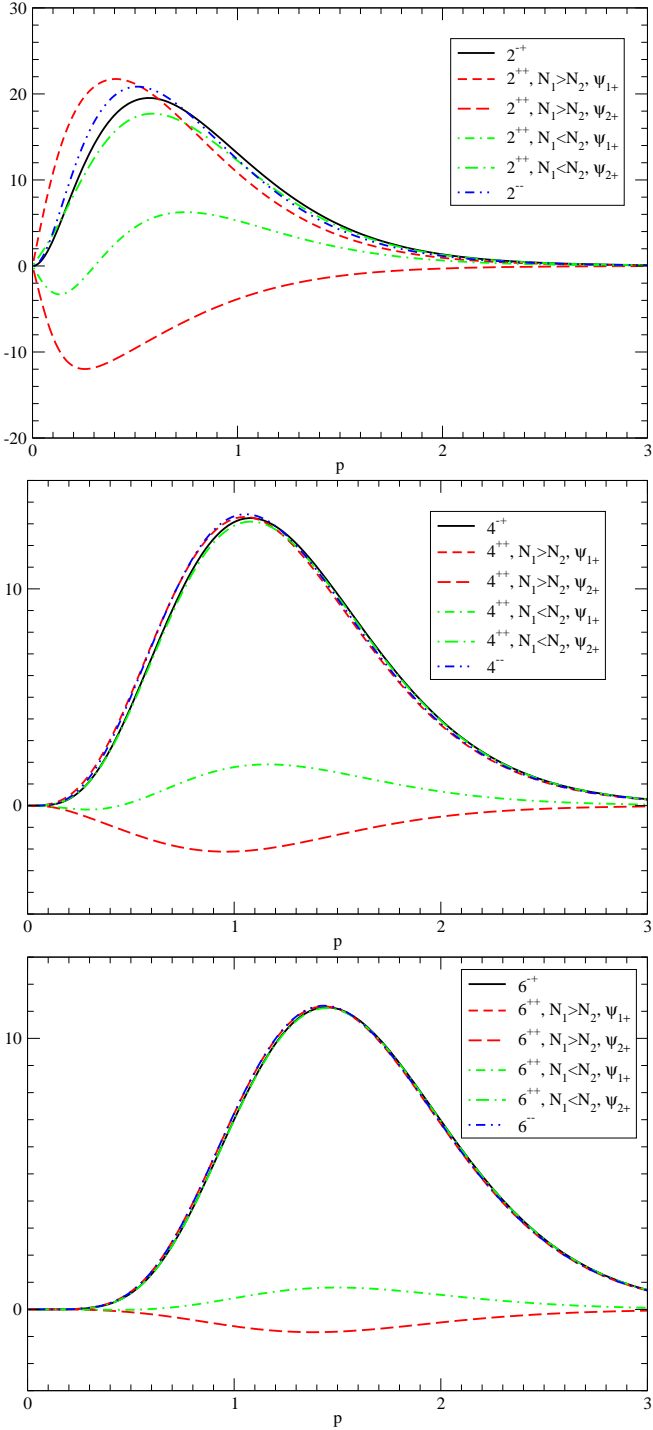

\includegraphics[width=\hsize,clip=]{J_2_wf_+.eps}
%
%
\includegraphics[width=\hsize,clip=]{J_4_wf_+.eps}
%
%
\includegraphics[width=\hsize,clip=]{J_6_wf_+.eps}
\caption{Components $\psi_+(p)$ of mesons with $J=2,4,6$ and $n=0$.
All quantities are given in appropriate units of $\sqrt{\sigma}$.}
\label{fig:J_246_wf}
\end{figure}

The radial wave functions of states with $n=0$ and increasing $J$
are shown in Fig. \ref{fig:J_246_wf}  for $J=2,4,6$.
Already for
$J=2$ four of these functions look rather similar. These are
the ones for the mesons of category 1 and 3 and in each case the larger function
of the two mesons of category 2, which is in the first case $\psi_{1+}(p)$ and
in the second case $\psi_{2+}(p)$. The two other functions ($\psi_{2+}(p)$
and $\psi_{1+}(p)$ in the first and second case, respectively) are smaller.
From Fig. \ref{fig:mixing} one sees that ${\cal N}_1$ (or ${\cal N}_2$)
are both larger than $0.8$ in the first and second case, respectively.
For $J=4$ and even more for $J=6$ the mixing becomes very small
and consequently the physics is determined by the
four almost indistinguishable radial functions. Consequently there appears
approximate degeneracy of all mesons for given $n$ at large $J$.
This is in accord with the
discussion in the preceeding section.
Obviously the wave function for larger $J$ are strongly suppressed at 
small momenta,
i.e., the radial function goes to zero as $p^{J}$ for mesons of categories 1 
and 3
and as $p^{J-1}$ for mesons of category 2. It is this feature which
explains a fast chiral restoration with increasing $J$, because the
chiral symmetry breaking dynamical mass $M(p)$ is essential only
at small momenta.

\begin{figure}
\includegraphics[width=\hsize,clip=]{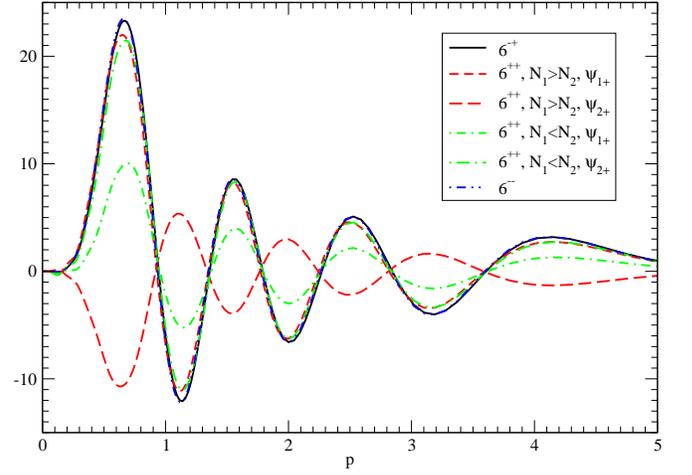}
\caption{Components $\psi_+(p)$ of mesons with $J=6$ and $n=6$.
All quantities are given in appropriate units of $\sqrt{\sigma}$.}
\label{fig:J_6_6}
\end{figure}

For higher radial excitations the wave functions
acquire nodes and the first bump is shifted towards lower momenta 
as we show in
Fig. \ref{fig:J_6_6} for $J=6$ and $n=6$. 
That explains
why at given $J$ for increasing $n$ the mixing becomes larger.

\section{Conclusions}

In this paper we have demonstrated explicitly a fast effective chiral symmetry
restoration in excited mesons with increasing $J$ and a slow restoration
with increasing $n$. To this end we used a model where
the only interaction between quarks is linear instantaneous Coulomb-like
confining interaction between color charges. This model contains all the
necessary ingredients - it is confining and manifestly chirally symmetric.
Chiral symmetry breaking is provided in the standard way through the
nonperturbative self interactions of quarks while the meson spectrum
is obtained from the Bethe-Salpeter equation. 

The key implication of
dynamical chiral symmetry breaking is that the quarks acquire the
dynamical Lorentz-scalar mass which is exclusively of quantum (loop)
origin. This dynamical mass is strongly momentum-dependent and vanishes
fast at larger momenta. In highly excited states the quantum (loop)
effects must be suppressed relative the classical
contributions and the spectrum exhibits an approximate
$SU(2)_L \times SU(2)_R$ and $U(1)_A$ symmetry. Microscopically this is
because a typical momentum of valence quarks increases higher in the
spectrum and consequently their dynamical Lorentz-scalar mass, which
violates chiral symmetry, becomes small and asymptotically vanishes.

Within  nonrelativistic or semirelativistic quark models the
effective chiral restoration cannot be reproduced as a matter of principle,
because (i) the constituent mass of quarks is an input parameter and
is momentum-independent, (ii) the notion of the quark chirality is
absent within a nonrelativistic or semirelativistic approach, (iii) a
Lorentz-scalar confining interaction, which is typically assumed in these
quark models, breaks chiral symmetry manifestly.

We have demonstrated explicitly that in the high-lying mesons all higher Fock
components are strongly suppressed relative the leading $\bar q q$
one and
 a very fast restoration of chiral symmetry happens with increasing
$J$ and rather slow one with increasing $n$. A reason for the former is that
with larger $J$ the radial wave
function vanishes at small relative momenta, and hence the large 
chiral-symmetry breaking dynamical mass becomes irrelevant. The slower
rate of the restoration in the latter case is due to the fact that at $J=0$
the wave function at small momenta does not vanish for pseudoscalar
mesons, so that even though a typical momentum of quarks increases with
$n$, the small momentum dynamical mass of quarks always contributes to
some extent.

All possible states for a given $n$ and large $J$ fall into higher
representations $[(0,1/2) \oplus (1/2,0)] \times [(0,1/2) \oplus (1/2,0)]$
that combine all possible chiral multiplets with the given $J$ and $n$.
The same property is observed for large $n$ at a given $J$, though a degree
of degeneracy is smaller. This property means that in both these cases
the quantum loop effects become irrelevant and all possible states with
different orientations of quark chirality become equivalent. 

We have shown that both radial and angular Regge trajectories are
linear asymptotically. At smaller $J$ and/or $n$ the linear form is
broken due to chiral symmetry breaking effects. Each Regge trajectory
is characterized by the proper chiral index and hence the amount of independent
Regge trajectories coincides with the amount of chiral representations.
Consequently for many mesons there are two independent Regge trajectories at
given $I,J^{PC}$, because in the chirally restored regime there are two
independent and degenerate states with the same $I,J^{PC}$ which belong to
different chiral representations.

\section{Acknowledgments}

RFW acknowledges helpful discussions with R. Alkofer, M. Kloker,
and A. Krassnigg. 
This work was supported by the Austrian Science Fund (projects
P16310-N08  and P19168-N16) and by DFG (project Al 279/5-1).

\section{Appendix A}
In this Appendix we discuss a general parametrization of the
vertex function and its specific form for mesons of all
categories with the instantaneous interaction.

For a meson with mass $\mu$, spin $J$,
intrinsic parity $P$ and $C$-parity $C$
the vertex function $\chi^{PC}_{JM}(P,p)$
can be written as a sum of eight (four for $J=0$)
components, where $M=-J,\ldots,J$ is the $z$-projection of $J$ in the rest
frame of the meson. $P^\nu$ and $p^\nu$ are the total
and relative four-momenta of the quark-antiquark system, respectively.
Each component contains a function $\tilde{\chi}_i$ which depends on the two invariants   
$p^2$ and $(p.P)^2$ (notice that $p.P$ is Lorentz-invariant but changes its
sign under the $C$-parity transformation).
Since we are only interested in spectra
in this paper we can restrict ourselves to the rest frame
of the meson (i.e., the frame with $P^\nu=(\mu,0,0,0)$),
where the vertex function depends on $\mu$, $p_0$ and $\vec{p}$
and each component contains a function $\chi_i$ of the two variables
$p_0^2$ and $|\vec{p}|$.
Some of the components contain a factor $p_0$ and are thus odd functions of $p_0$.
At the same time, for an instantaneous interaction the vertex function
in the rest frame is independent of $p_0$.
Thus all components which are odd
in $p_0$ must vanish. The 
function $\chi_i$ of the remaining components depend only on $p=|\vec{p}|$.
Depending on the quantum numbers $J^{PC}$ there are
four different categories of mesons.
In the following we give the general expressions for the vertex function
$\chi^{PC}_{JM}(P,p)$ in a general frame, 
$\chi^{PC}_{JM}(\mu,p_0,\vec{p})$ in the rest frame and
$\chi^{PC}_{JM}(\mu,\vec{p})$
in the rest frame and an instantaneous interaction.

Category 1: Mesons with $J^{-+}$ for $J=2n$ and $J^{+-}$ for $J=2n+1$, respectively
have the vertex function 
\begin{equation}
\begin{array}{rcl}\chi^{PC}_{JM}(P,p)&=&
                \gamma_5(\epsilon^{(J,M)}(P)\cdot p)\tilde{\chi}_1\\ &+&
		i\gamma_\mu\epsilon^{\mu\nu\rho\sigma}(\epsilon^{(J,M)}_\nu(P)\cdot p)
		P_\rho p_\sigma(p\cdot P)\tilde{\chi}_2\\ &+&
                \gamma_5\gamma^\mu
		P_\mu(\epsilon^{(J,M)}(P)\cdot p)\tilde{\chi}_3\\ &+&
		\gamma_5\gamma^\mu p_\mu(\epsilon^{(J,M)}(P)\cdot p)(p\cdot P)\tilde{\chi}_4\\ &+&
		\gamma_5\gamma^\mu(\epsilon^{(J,M)}_\mu(P)\cdot p)(p\cdot P)\tilde{\chi}_5
		\\ &+&
		i\gamma_5\sigma^{\mu\nu}
		P_\mu p_\nu(\epsilon^{(J,M)}(P)\cdot p)\tilde{\chi}_6\\ &+&
		i\gamma_5\sigma^{\mu\nu}P_\mu(\epsilon^{(J,M)}_\nu(P)\cdot p)\tilde{\chi}_7\\ &+&
		i\gamma_5\sigma^{\mu\nu}p_\mu(\epsilon^{(J,M)}_\nu(P)\cdot p)(p\cdot P)\tilde{\chi}_8
		.
		\end{array}
\label{eq:bsatype1}
\end{equation}
Here
\begin{equation}
(\epsilon^{(J,M)}(P)\cdot p)=
\epsilon^{(J,M)}_{\mu_1\ldots\mu_J}(P)p^{\mu_1}\ldots p^{\mu_J},
\end{equation}
\begin{equation}
(\epsilon^{(J,M)}_\mu(P)\cdot p)=
\epsilon^{(J,M)}_{\mu\mu_2\ldots\mu_J}(P)p^{\mu_2}\ldots p^{\mu_J},
\end{equation}
and $\epsilon^{(J,M)}_{\mu_1\ldots\mu_J}(P)$ are
the polarization tensors of rank $J$. They have the properties
\begin{equation}
\epsilon^{(J,M)}_{\ldots \mu_i\ldots \mu_j\ldots}(P)=
\epsilon^{(J,M)}_{\ldots \mu_j\ldots \mu_i\ldots}(P),
\end{equation}
\begin{equation}
g^{\mu_i\mu_j}\epsilon^{(J,M)}_{\ldots \mu_i\ldots \mu_j \ldots}(P)=0,
\end{equation}
and
\begin{equation}
\epsilon^{(J,M)}_{\ldots \mu\ldots}(P)P^\mu=0.
\end{equation}
Contracting a polarization tensor with $J$ vectors $a_i$, $i=1,\ldots J$
yields in
the rest frame
\begin{equation}
\begin{array}{l}
\epsilon^{(J,M)}_{\mu_1\ldots \mu_J}((\mu,0,0,0))a_1^{\mu_1}\ldots a_J^{\mu_J}=\\
\sqrt{\frac{(2J-1)!!}{J!}}
\{\{\ldots\{\vec{a}_1\otimes\vec{a}_2\}_2\otimes\ldots\}_{J-1}\otimes\vec{a}_J\}_{JM},
\end{array}
\end{equation}
where $\{J_1\otimes J_2\}_{JM}$ is the usual coupling of two spherical tensors
of rank $J_1$ and $J_2$ to a spherical tensor of rank $J$.
Thus the vertex function (\ref{eq:bsatype1}) can be written in the form
\begin{equation}
\begin{array}{rcl}\chi^{PC}_{JM}(\mu,p_0,\vec{p})&=&\gamma_5Y_{JM}(\hat{p})\chi_1(p_0^2,|\vec{p}|)\\&+&
                \mu\gamma_0\gamma_5Y_{JM}(\hat{p})\chi_2(p_0^2,|\vec{p}|)\\ &+&
		\mu\gamma_0\gamma_5\left\{Y_{J+1}(\hat{p})\otimes \vec{\gamma}\right\}_{JM}\chi_3(p_0^2,|\vec{p}|)\\ &+&
		\mu\gamma_0\gamma_5\left\{Y_{J-1}(\hat{p})\otimes\vec{\gamma}\right\}_{JM}\chi_4(p_0^2,|\vec{p}|)\\ &+&
                p_0\left\{Y_{J}(\hat{p})\otimes \vec{\gamma}\right\}_{JM}\chi_5(p_0^2,|\vec{p}|)\\ &+&
                p_0\mu\gamma_0\left\{Y_{J}(\hat{p})\otimes \vec{\gamma}\right\}_{JM}\chi_6(p_0^2,|\vec{p}|)\\ &+&
		p_0\mu\gamma_5\left\{Y_{J+1}(\hat{p})\otimes \vec{\gamma}\right\}_{JM}\chi_7(p_0^2,|\vec{p}|)\\ &+&
	        p_0\mu\gamma_5\left\{Y_{J-1}(\hat{p})\otimes\vec{\gamma}\right\}_{JM}\chi_8(p_0^2,|\vec{p}|),
		\end{array}
\end{equation}
where the $Y_{JM}(\hat{p})$ are the spherical harmonics.
For $J=0$ the components $4,5,6,8$ are absent.
The components 1--4 are even and the components 5--8 odd functions of $p_0$, respectively.
For an instantaneous interaction we are left with the vertex function
\begin{equation}
\begin{array}{rcl}\chi^{PC}_{JM}(m,\vec{p})&=&\gamma_5Y_{JM}(\hat{p})\chi_1(p)+
                \mu\gamma_0\gamma_5Y_{JM}(\hat{p})\chi_2(p)\\ &+&
		\mu\gamma_0\gamma_5\left\{Y_{J+1}(\hat{p})\otimes \vec{\gamma}\right\}_{JM}\chi_3(p)\\
		&+&
	        \mu\gamma_0\gamma_5\left\{Y_{J-1}(\hat{p})\otimes\vec{\gamma}\right\}_{JM}\chi_4(p)
		\end{array}
\end{equation}
for mesons of this category.

Category 2: Mesons with $J^{++}$ for $J=2n$ and $J^{--}$ for $J=2n+1$, respectively
have the vertex function 
\begin{equation}
\label{eq:bsatype2}
\begin{array}{rcl}\chi^{PC}_{JM}(P,p)&=&
                (\epsilon^{(J,M)}(P)\cdot p)\tilde{\chi}_1\\ &+&
		i\gamma_5\gamma_\mu\epsilon^{\mu\nu\rho\sigma}(\epsilon^{(J,M)}_\nu(P)\cdot p)
		P_\rho p_\sigma\tilde{\chi}_2\\ &+&
                \gamma^\mu
		P_\mu(\epsilon^{(J,M)}(P)\cdot p)(p\cdot P)\tilde{\chi}_3\\ &+&
		\gamma^\mu p_\mu(\epsilon^{(J,M)}(P)\cdot p)\tilde{\chi}_4\\ &+&
		\gamma^\mu(\epsilon^{(J,M)}_\mu(P)\cdot p)\tilde{\chi}_5
		\\ &+&
		i\sigma^{\mu\nu}
		P_\mu p_\nu(\epsilon^{(J,M)}(P)\cdot p)\tilde{\chi}_6\\ &+&
		i\sigma^{\mu\nu}P_\mu(\epsilon^{(J,M)}_\nu(P)\cdot p)\tilde{\chi}_7\\ &+&
		i\sigma^{\mu\nu}Pp_\mu(\epsilon^{(J,M)}_\nu(P)\cdot p)(p\cdot P)\tilde{\chi}_8
		.
		\end{array}
\end{equation}
In the rest frame it can be written as
\begin{equation}
\begin{array}{rcl}\chi_{JM}^{PC}(\mu,p_0,\vec{p})&=&Y_{JM}(\hat{p})\chi_1(p_0^2,|\vec{p}|)\\ &+&
		\left\{Y_{J+1}(\hat{p})\otimes \vec{\gamma}\right\}_{JM}\chi_2(p_0^2,|\vec{p}|)\\ &+&
		\left\{Y_{J-1}(\hat{p})\otimes \vec{\gamma}\right\}_{JM}\chi_3(p_0^2,|\vec{p}|)\\ &+&
                \mu\gamma_5\left\{Y_{J}(\hat{p})\otimes \vec{\gamma}\right\}_{JM}\chi_4(p_0^2,|\vec{p}|)\\ &+&
		\mu\gamma_0\left\{Y_{J+1}(\hat{p})\otimes \vec{\gamma}\right\}_{JM}\chi_5(p_0^2,|\vec{p}|)\\ &+&
		\mu\gamma_0\left\{Y_{J-1}(\hat{p})\otimes \vec{\gamma}\right\}_{JM}\chi_6(p_0^2,|\vec{p}|)\\ &+&
		p_0\gamma_0Y_{JM}(\hat{p})\chi_7(p_0^2,|\vec{p}|)\\ &+&
		p_0\mu\gamma_0\gamma_5\left\{Y_{J}(\hat{p})\otimes \vec{\gamma}\right\}_{JM}\chi_8(p_0^2,|\vec{p}|).
		\end{array}
\end{equation}
For $J=0$ the components $3,4,6,8$ are absent.
The components 1--6 are even and the components 7,8 odd functions of $p_0$, respectively.
For an instantaneous interaction we are left with the vertex function
\begin{equation}
\begin{array}{rcl}\chi_{JM}^{PC}(\mu,\vec{p})&=&Y_{JM}(\hat{p})\chi_1(p)+
		\left\{Y_{J+1}(\hat{p})\otimes \vec{\gamma}\right\}_{JM}\chi_2(p)\\ &+&
		\left\{Y_{J-1}(\hat{p})\otimes \vec{\gamma}\right\}_{JM}\chi_3(p)\\ &+&
                \mu\gamma_5\left\{Y_{J}(\hat{p})\otimes \vec{\gamma}\right\}_{JM}\chi_4(p)\\ &+&
		\mu\gamma_0\left\{Y_{J+1}(\hat{p})\otimes \vec{\gamma}\right\}_{JM}\chi_5(p)\\ &+&
		\mu\gamma_0\left\{Y_{J-1}(\hat{p})\otimes \vec{\gamma}\right\}_{JM}\chi_6(p),
		\end{array}
\end{equation}
for mesons of this category.

Category 3: Mesons with $J^{--}$ for $J=2n$ and $J^{++}$ for $J=2n+1$, respectively
have the vertex function 
\begin{equation}
\label{eq:bsatype3}
\begin{array}{rcl}\chi^{PC}_{JM}(p,P)&=&
                \gamma_5(\epsilon^{(J,M)}(P)\cdot p)(p\cdot P)\tilde{\chi}_1\\ &+&
		i\gamma_\mu\epsilon^{\mu\nu\rho\sigma}(\epsilon^{(J,M)}_\nu(P)\cdot p)
		P_\rho p_\sigma\tilde{\chi}_2\\ &+&
                \gamma_5\gamma^\mu
		P_\mu(\epsilon^{(J,M)}(P)\cdot p)(p\cdot P)\tilde{\chi}_3\\ &+&
		\gamma_5\gamma^\mu p_\mu(\epsilon^{(J,M)}(P)\cdot p)\tilde{\chi}_4\\ &+&
		\gamma_5\gamma^\mu(\epsilon^{(J,M)}_\mu(P)\cdot p)\tilde{\chi}_5
		\\ &+&
		i\gamma_5\sigma^{\mu\nu}
		P_\mu p_\nu(\epsilon^{(J,M)}(P)\cdot p)(p\cdot P)\tilde{\chi}_6\\ &+&
		i\gamma_5\sigma^{\mu\nu}P_\mu(\epsilon^{(J,M)}_\nu(P)\cdot p)(p\cdot
		P)\tilde{\chi}_7\\ &+&
		i\gamma_5\sigma^{\mu\nu}p_\mu(\epsilon^{(J,M)}_\nu(P)\cdot p)\tilde{\chi}_8
		.
		\end{array}
\end{equation}
In the rest frame it can be written as
\begin{equation}
\begin{array}{rcl}\chi_{JM}^{PC}(\mu,p_0,\vec{p})&=&
                \mu\left\{Y_{J}(\hat{p})\otimes \vec{\gamma}\right\}_{JM}\chi_1(p_0^2,|\vec{p}|)\\ &+&
		\gamma_5\left\{Y_{J+1}(\hat{p})\otimes \vec{\gamma}\right\}_{JM}\chi_2(p_0^2,|\vec{p}|)\\ &+&
		\gamma_5\left\{Y_{J-1}(\hat{p})\otimes \vec{\gamma}\right\}_{JM}\chi_3(p_0^2,|\vec{p}|)\\ &+&
                \gamma_0\left\{Y_{J}(\hat{p})\otimes \vec{\gamma}\right\}_{JM}\chi_4(p_0^2,|\vec{p}|)\\ &+&
                p_0\mu\gamma_5 Y_{JM}\chi_5(p_0^2,|\vec{p}|)\\ &+&
                p_0\gamma_0\gamma_5 Y_{JM}\chi_6(p_0^2,|\vec{p}|)\\ &+&
		p_0\gamma_0\gamma_5\left\{Y_{J+1}(\hat{p})\otimes\vec{\gamma}\right\}_{JM}\chi_7(p_0^2,|\vec{p}|)\\ &+&
		p_0\gamma_0\gamma_5\left\{Y_{J-1}(\hat{p})\otimes \vec{\gamma}\right\}_{JM}\chi_8(p_0^2,|\vec{p}|)
		\end{array}
\end{equation}
For $J=0$ the components $1,3,4,8$ are absent.
The components 1--4 are even and the components 5--8 odd functions of $p_0$, respectively.
For an instantaneous interaction we are left with the vertex function
\begin{equation}
\begin{array}{rcl}\chi_{JM}^{PC}(\mu,\vec{p})&=&\mu\left\{Y_{J}(\hat{p})\otimes
\vec{\gamma}\right\}_{JM}\chi_1(p)\\ &+&
		\gamma_5\left\{Y_{J+1}(\hat{p})\otimes \vec{\gamma}\right\}_{JM}\chi_2(p)\\ &+&
		\gamma_5\left\{Y_{J-1}(\hat{p})\otimes \vec{\gamma}\right\}_{JM}\chi_3(p)\\ &+&
                \gamma_0\left\{Y_{J}(\hat{p})\otimes \vec{\gamma}\right\}_{JM}\chi_4(p).
		\end{array}
\end{equation}
for mesons of this category.

Category 4: Mesons with $J^{+-}$ for $J=2n$ and $J^{-+}$ for $J=2n+1$, respectively
have the vertex function 
\begin{equation}
\label{eq:bsatype4}
\begin{array}{rcl}\chi^{PC}_{JM}(p,P)&=&
                (\epsilon^{(J,M)}(P)\cdot p)(p\cdot P)\tilde{\chi}_1\\ &+&
		i\gamma_5\gamma_\mu\epsilon^{\mu\nu\rho\sigma}(\epsilon^{(J,M)}_\nu(P)\cdot p)
		P_\rho p_\sigma(p\cdot P)\tilde{\chi}_2\\ &+&
                \gamma^\mu
		P_\mu(\epsilon^{(J,M)}(P)\cdot p)\tilde{\chi}_3\\ &+&
		\gamma^\mu p_\mu(\epsilon^{(J,M)}(P)\cdot p)(p\cdot P)\tilde{\chi}_4\\ &+&
		\gamma^\mu(\epsilon^{(J,M)}_\mu(P)\cdot p)(p\cdot P)\tilde{\chi}_5
		\\ &+&
		i\sigma^{\mu\nu}
		P_\mu p_\nu(\epsilon^{(J,M)}(P)\cdot p)(p\cdot P)\tilde{\chi}_6\\ &+&
		i\sigma^{\mu\nu}P_\mu(\epsilon^{(J,M)}_\nu(P)\cdot p)(p\cdot P)\tilde{\chi}_7\\ &+&
		i\sigma^{\mu\nu}p_\mu(\epsilon^{(J,M)}_\nu(P)\cdot p)\tilde{\chi}_8
		.
		\end{array}
\end{equation}
In the rest frame it can be written as
\begin{equation}
\begin{array}{rcl}\chi_{JM}^{PC}(\mu,p_0,\vec{p})&=&
		\mu\gamma_0Y_{JM}(\hat{p})\chi_1(p_0^2,|\vec{p}|)\\ &+&
		\gamma_0\gamma_5\left\{Y_{J}(\hat{p})\otimes \vec{\gamma}\right\}_{JM}\chi_2(p_0^2,|\vec{p}|)\\ &+&
                p_0\mu Y_{JM}(\hat{p})\chi_3(p_0^2,|\vec{p}|)\\ &+&
		p_0\mu\left\{Y_{J+1}(\hat{p})\otimes \vec{\gamma}\right\}_{JM}\chi_4(p_0^2,|\vec{p}|)\\ &+&
		p_0\mu\left\{Y_{J-1}(\hat{p})\otimes \vec{\gamma}\right\}_{JM}\chi_5(p_0^2,|\vec{p}|)\\ &+&
                p_0\gamma_5\left\{Y_{J}(\hat{p})\otimes \vec{\gamma}\right\}_{JM}\chi_6(p_0^2,|\vec{p}|)\\ &+&
		p_0\gamma_0\left\{Y_{J+1}(\hat{p})\otimes \vec{\gamma}\right\}_{JM}\chi_7(p_0^2,|\vec{p}|)\\ &+&
		p_0\gamma_0\left\{Y_{J-1}(\hat{p})\otimes \vec{\gamma}\right\}_{JM}\chi_8(p_0^2,|\vec{p}|)
		\end{array}
\end{equation}
For $J=0$ the components $2,5,6,8$ are absent.
The components 1,2 are even and the components 3--8 odd functions of $p_0$, respectively.
For an instantaneous interaction we are left with the vertex function
\begin{equation}
\begin{array}{rcl}\chi_{JM}^{PC}(\mu,\vec{p})&=&
		\mu\gamma_0Y_{JM}(\hat{p})\chi_1(p)\\ &+&
		\gamma_0\gamma_5\left\{Y_{J}(\hat{p})\otimes \vec{\gamma}\right\}_{JM}\chi_2(p)
		\end{array}
\end{equation}
for mesons of this category.

\section{Appendix B}
Given the parametrizations of the vertex function in Appendix A, we
derive in the present Appendix from the general Bethe-Salpeter
equation systems of coupled equations for all possible mesons categories. 

\bigskip
\bigskip
\bigskip

With the respective ansatz for the vertex function as given in the previous section
the BSE for a meson of given quantum numbers 
becomes an equation between two Dirac-matrices. By projecting out the functions
$\chi_i(p)$ on the left hand side one arrives at a system of coupled integral
equations. The projection can be performed by applying the standard rules of traces of
Dirac matrices and with help of 
%
%
\begin{equation}
\frac{4\pi}{2J+1}Y_J(\hat{p})\cdot Y_J(\hat{q})=P_J(\hat{p}\cdot\hat{q}),
\end{equation}
\begin{widetext}
\begin{equation}
\frac{4\pi}{2J+1}\sum\limits_{k=1}^3
\{ Y_{J_1}(\hat{p})\otimes \vec{e}_k\}_J\cdot \{ Y_{J_2}(\hat{q})\otimes \vec{e}_k\}_J=
(-1)^{J+J_1+1}\delta_{J_1J_2}P_{J_1}(\hat{p}\cdot\hat{q}),
\end{equation}
\begin{equation}
\frac{4\pi}{2J+1}Y_J(\hat{p})\cdot\{ Y_{J_2}(\hat{q})\otimes \hat{q}\}_J=
\sqrt{\frac{2J_2+1}{2J+1}}C_{J_2010}^{J0}P_J(\hat{p}\cdot\hat{q})=
\left\{\begin{array}{rl}
\displaystyle\sqrt{\frac{J}{2J+1}}P_J(\hat{p}\cdot\hat{q}),&J_2=J-1\\
\displaystyle-\sqrt{\frac{J+1}{2J+1}}P_J(\hat{p}\cdot\hat{q}),&J_2=J+1\\
\displaystyle 0,&\rm else,
\end{array}\right.
\end{equation}
\begin{equation}
\frac{4\pi}{2J+1}\{ Y_{J_1}(\hat{p})\otimes \hat{q}\}_J\cdot Y_J(\hat{q})=
(-1)^{J+J_1}C_{10J0}^{J_10}P_{J_1}(\hat{p}\cdot\hat{q})=
\left\{\begin{array}{rl}
\displaystyle\sqrt{\frac{J}{2J+1}}P_{J-1}(\hat{p}\cdot\hat{q}),&J_1=J-1\\
\displaystyle-\sqrt{\frac{J+1}{2J+1}}P_{J+1}(\hat{p}\cdot\hat{q}),&J_1=J+1\\
\displaystyle 0,&\rm else,
\end{array}\right.
\end{equation}
%
\begin{eqnarray}
\displaystyle
\frac{4\pi}{2J+1}\{ Y_{J_1}(\hat{p})\otimes \hat{q}\}_J\cdot 
\{ Y_{J_2}(\hat{q})\otimes \hat{q}\}_J&=&
(-1)^{J+J_1}\sqrt{\frac{2J_2+1}{2J+1}}C_{10J0}^{J_10}C_{J_2010}^{J0}P_{J_1}(\hat{p}\cdot\hat{q})
\nonumber \\&=&
\left\{\begin{array}{rl}
\displaystyle\frac{J}{2J+1}P_{J-1}(\hat{p}\cdot\hat{q}),&J_1=J_2=J-1\\
\displaystyle-\frac{\sqrt{J(J+1)}}{2J+1}P_{J-1}(\hat{p}\cdot\hat{q}),&J_1=J-1,J_2=J+1\\
\displaystyle-\frac{\sqrt{J(J+1)}}{2J+1}P_{J+1}(\hat{p}\cdot\hat{q}),&J_1=J+1,J_2=J-1\\
\displaystyle\frac{J+1}{2J+1}P_{J+1}(\hat{p}\cdot\hat{q}),&J_1=J_2=J+1\\
\displaystyle 0,&\rm else,
\end{array}\right.
\end{eqnarray}
%
\begin{eqnarray}
\frac{4\pi}{2J+1}i\sum\limits_{j,k,l=1}^3\epsilon_{jkl}
\{ Y_{J_1}(\hat{p})\otimes \vec{e}_j\}_J\cdot \{ Y_{J_2}(\hat{q})\otimes \vec{e}_k\}_J
\frac{q_l}{q}&=&
(-1)^{J_1+J_2+1}\sqrt{6(2J_2+1)}C_{J_2010}^{J_10}
\left\{\begin{array}{lll}J_1&1&J\\ 1&J_2&1\end{array}\right\}
P_{J_1}(\hat{p}\cdot\hat{q})
\nonumber \\&=&
\left\{\begin{array}{rl}
\displaystyle\sqrt{\frac{J+1}{2J+1}}P_{J-1}(\hat{p}\cdot\hat{q}),&J_1=J-1,J_2=J\\
\displaystyle-\sqrt{\frac{J+1}{2J+1}}P_{J}(\hat{p}\cdot\hat{q}),&J_1=J,J_2=J-1\\
\displaystyle-\sqrt{\frac{J}{2J+1}}P_{J}(\hat{p}\cdot\hat{q}),&J_1=J,J_2=J+1\\
\displaystyle\sqrt{\frac{J}{2J+1}}P_{J+1}(\hat{p}\cdot\hat{q}),&J_1=J+1,J_2=J\\
\displaystyle 0,&\rm else.
\end{array}\right.
\end{eqnarray}
\end{widetext}

For mesons of category 1 the BSE becomes a coupled system of four integral equations
%
\begin{widetext}
\begin{subequations}
\label{eq:fullbsetype1}
\begin{eqnarray}
\chi_1(p)&=&\frac{1}{2}\int \frac{d^3q}{(2\pi)^3} \frac{V(k)}
{\omega^2(q)-\frac{\mu^2}{4}}P_J(\hat{p}\cdot\hat{q})\nonumber \\&\times&
\left\{\omega(q)\chi_1(q)+
\frac{\mu^2}{2\bar{\omega}(q)}
\left[M(q)\chi_2(q)-
q\left(\sqrt{\frac{J+1}{2J+1}}\chi_3(q)-
\sqrt{\frac{J}{2J+1}}\chi_4(q)\right)\right]\right\}
,
\label{eq:fullbsetype1a}
\end{eqnarray}
\begin{eqnarray}
\chi_2(p)&=&\frac{1}{2}\int \frac{d^3q}{(2\pi)^3} \frac{V(k)}
{\omega^2(q)-\frac{\mu^2}{4}}P_J(\hat{p}\cdot\hat{q})\nonumber \\&\times&
\frac{M(q)}{\bar{\omega}(q)}
\left\{
\frac{1}{2}\chi_1(q)+
\frac{\omega(q)}{\bar{\omega}(q)}\left[M(q)\chi_2(q)-
q\left(\sqrt{\frac{J+1}{2J+1}}\chi_3(q)-
\sqrt{\frac{J}{2J+1}}\chi_4(q)\right)\right]\right\}
,
\label{eq:fullbsetype1b}
\end{eqnarray}
\begin{eqnarray}
\chi_3(p)&=&-\sqrt{\frac{J+1}{2J+1}}\frac{1}{2}\int \frac{d^3q}{(2\pi)^3} 
\frac{V(k)}{\omega^2(q)-\frac{\mu^2}{4}}P_{J+1}(\hat{p}\cdot\hat{q})
\nonumber \\&\times&
\frac{q}{\bar{\omega}(q)}
\left\{
\frac{1}{2}\chi_1(q)+
\frac{\omega(q)}{\bar{\omega}(q)}\left[M(q)\chi_2(q)-
q\left(\sqrt{\frac{J+1}{2J+1}}\chi_3(q)-
\sqrt{\frac{J}{2J+1}}\chi_4(q)\right)\right]\right\}
,
\label{eq:fullbsetype1c}
\end{eqnarray}
\begin{eqnarray}
\chi_4(p)&=&\sqrt{\frac{J}{2J+1}}\frac{1}{2}\int \frac{d^3q}{(2\pi)^3} 
\frac{V(k)}{\omega^2(q)-\frac{\mu^2}{4}}P_{J-1}(\hat{p}\cdot\hat{q})
\nonumber \\&\times&
\frac{q}{\bar{\omega}(q)}
\left\{
\frac{1}{2}\chi_1(q)+
\frac{\omega(q)}{\bar{\omega}(q)}\left[M(q)\chi_2(q)-
q\left(\sqrt{\frac{J+1}{2J+1}}\chi_3(q)-
\sqrt{\frac{J}{2J+1}}\chi_4(q)\right)\right]\right\}
.
\label{eq:fullbsetype1d}
\end{eqnarray}
\end{subequations}
Obviously in the integrands there appear only two linear combinations
$\chi_1(q)$ and the term between the square brackets.
Thus one can reduce (\ref{eq:fullbsetype1}) to
a coupled system of two integral equations
\begin{subequations}
\label{eq:bsetype1}
\begin{eqnarray}
\omega(p)h(p)&=&\frac{1}{2}\int \frac{d^3q}{(2\pi)^3} V(k)
P_J(\hat{p}\cdot\hat{q})
\left(h(q)+\frac{\mu^2}{4\omega(q)}g(q)\right),
%
\label{eq:bsetype1a}
\\
\left(\omega(p)-\frac{\mu^2}{4\omega(p)}\right)g(p)&=&h(p)
\nonumber \\&+&\frac{1}{2}\int \frac{d^3q}{(2\pi)^3} V(k)
\frac{M(p)M(q)P_J(\hat{p}\cdot\hat{q})+pq\left(
\frac{J+1}{2J+1}P_{J+1}(\hat{p}\cdot\hat{q})
+\frac{J}{2J+1}P_{J-1}(\hat{p}\cdot\hat{q})\right)}{\bar{\omega}(p)\bar{\omega}(q)}
g(q)
\label{eq:bsetype1b}
\end{eqnarray}
\end{subequations}
for the two functions
\begin{equation}
h(p)=\frac{\chi_1(p)}{\omega(p)}
\end{equation}
and
\begin{equation}
g(p)=\frac{\omega(p)}{\omega^2(p)-\frac{\mu^2}{4}}\left[h(p)+2\frac{M(p)}{\bar{\omega}(p)}\chi_2(p)
-2\frac{p}{\bar{\omega}(p)}\left(\sqrt{\frac{J+1}{2J+1}}\chi_3(p)-
\sqrt{\frac{J}{2J+1}}\chi_4(p)\right)\right].
\end{equation}
\end{widetext}
For $J=0$ Eq. (\ref{eq:fullbsetype1d}) for $\chi_4$ is absent and all terms containing $\chi_4$
in the integrands of (\ref{eq:fullbsetype1a}--\ref{eq:fullbsetype1c}) vanish.

For mesons of category 2 the BSE becomes a coupled system of six  integral equations
\begin{widetext}
\begin{subequations}
\label{eq:fullbsetype2}
\begin{eqnarray}
\chi_1(p)&=&\frac{1}{2}\int \frac{d^3q}{(2\pi)^3} \frac{V(k)}{\omega^2(q)-\frac{\mu^2}{4}}P_J(\hat{p}\cdot\hat{q})
\frac{q}{\bar{\omega}(q)}\nonumber \\&\times&
\left\{
\frac{\omega(q)}{\bar{\omega}(q)}\left[q\chi_1(q)+
M(q)\left(\sqrt{\frac{J+1}{2J+1}}\chi_2(q)-
\sqrt{\frac{J}{2J+1}}\chi_3(q)\right)\right]
\right. \nonumber\\&&\left.
+\frac{\mu^2}{2}
\left[\sqrt{\frac{J+1}{2J+1}}\chi_5(q)-
\sqrt{\frac{J}{2J+1}}
\chi_6(q)\right]\right\},
\label{eq:fullbsetype2a}\
\end{eqnarray}
\begin{eqnarray}
\chi_2(p)&=&\frac{1}{2}\int \frac{d^3q}{(2\pi)^3} \frac{V(k)}{\omega^2(q)-\frac{\mu^2}{4}}P_{J+1}(\hat{p}\cdot\hat{q})
\nonumber \\&\times&
\left\{\omega(q)\sqrt{\frac{J}{2J+1}}\left[\sqrt{\frac{J}{2J+1}}\chi_2(q)+
\sqrt{\frac{J+1}{2J+1}}\chi_3(q)\right]
\right.\nonumber \\&&
+\frac{\omega(q)M(q)}{\bar{\omega}^2(q)}\sqrt{\frac{J+1}{2J+1}}\left[q\chi_1(q)+M(q)
\left(\sqrt{\frac{J+1}{2J+1}}\chi_2(q)-
\sqrt{\frac{J}{2J+1}}\chi_3(q)\right)\right]
 \nonumber\\&&
+\frac{\mu^2}{2\bar{\omega}(q)}\sqrt{\frac{J}{2J+1}}
\left[q\chi_4(q)+M(q)\left(\sqrt{\frac{J}{2J+1}}\chi_5(q)+
\sqrt{\frac{J+1}{2J+1}}
\chi_6(q)\right)\right]\nonumber\\&&\left.+
\frac{\mu^2M(q)}{2\bar{\omega}(q)}\sqrt{\frac{J+1}{2J+1}}\left[\sqrt{\frac{J+1}{2J+1}}\chi_5(q)-
\sqrt{\frac{J}{2J+1}}
\chi_6(q)\right]
\right\},
\label{eq:fullbsetype2b}\
\end{eqnarray}
\begin{eqnarray}
\chi_3(p)&=&\frac{1}{2}\int \frac{d^3q}{(2\pi)^3} \frac{V(k)}{\omega^2(q)-\frac{\mu^2}{4}}P_{J-1}(\hat{p}\cdot\hat{q})
\nonumber \\&\times&
\left\{\omega(q)\sqrt{\frac{J+1}{2J+1}}\left[\sqrt{\frac{J}{2J+1}}\chi_2(q)+
\sqrt{\frac{J+1}{2J+1}}\chi_3(q)\right]
\right.\nonumber \\&&
-\frac{\omega(q)M(q)}{\bar{\omega}^2(q)}\sqrt{\frac{J}{2J+1}}\left[q\chi_1(q)+M(q)
\left(\sqrt{\frac{J+1}{2J+1}}\chi_2(q)-
\sqrt{\frac{J}{2J+1}}\chi_3(q)\right)\right]
 \nonumber\\&&
+\frac{\mu^2}{2\bar{\omega}(q)}\sqrt{\frac{J+1}{2J+1}}
\left[q\chi_4(q)+M(q)\left(\sqrt{\frac{J}{2J+1}}\chi_5(q)+
\sqrt{\frac{J+1}{2J+1}}
\chi_6(q)\right)\right]\nonumber\\&&\left.-
\frac{\mu^2M(q)}{2\bar{\omega}(q)}\sqrt{\frac{J}{2J+1}}\left[\sqrt{\frac{J+1}{2J+1}}\chi_5(q)-
\sqrt{\frac{J}{2J+1}}
\chi_6(q)\right]
\right\},
\label{eq:fullbsetype2c}\
\end{eqnarray}
\begin{eqnarray}
\chi_4(p)&=&\frac{1}{2}\int \frac{d^3q}{(2\pi)^3} \frac{V(k)}{\omega^2(q)-\frac{\mu^2}{4}}P_{J}(\hat{p}\cdot\hat{q})
\nonumber \\&\times&
\left\{\frac{q}{2\bar{\omega}(q)}\left[\sqrt{\frac{J}{2J+1}}\chi_2(q)+
\sqrt{\frac{J+1}{2J+1}}\chi_3(q)\right]
\right.\nonumber \\&&\left.
+\frac{q\omega(q)}{\bar{\omega}^2(q)}
\left[q\chi_4(q)+M(q)\left(\sqrt{\frac{J}{2J+1}}\chi_5(q)+
\sqrt{\frac{J+1}{2J+1}}
\chi_6(q)\right)\right]
\right\},
\label{eq:fullbsetype2d}
\end{eqnarray}
\begin{eqnarray}
\chi_5(p)&=&\frac{1}{2}\int \frac{d^3q}{(2\pi)^3} \frac{V(k)}{\omega^2(q)-\frac{\mu^2}{4}}P_{J+1}(\hat{p}\cdot\hat{q})
\nonumber \\&\times&
\left\{\frac{M(q)}{2\bar{\omega}(q)}\sqrt{\frac{J}{2J+1}}\left[\sqrt{\frac{J}{2J+1}}\chi_2(q)+
\sqrt{\frac{J+1}{2J+1}}\chi_3(q)\right]
\right.\nonumber \\&&
+\frac{1}{2\bar{\omega}(q)}\sqrt{\frac{J+1}{2J+1}}\left[q\chi_1(q)+M(q)
\left(\sqrt{\frac{J+1}{2J+1}}\chi_2(q)-
\sqrt{\frac{J}{2J+1}}\chi_3(q)\right)\right]
 \nonumber\\&&
+\frac{M(q)\omega(q)}{\bar{\omega}^2(q)}\sqrt{\frac{J}{2J+1}}
\left[q\chi_4(q)+M(q)\left(\sqrt{\frac{J}{2J+1}}\chi_5(q)+
\sqrt{\frac{J+1}{2J+1}}
\chi_6(q)\right)\right]\nonumber\\&&\left.+
\omega(q)\sqrt{\frac{J+1}{2J+1}}\left[\sqrt{\frac{J+1}{2J+1}}\chi_5(q)-
\sqrt{\frac{J}{2J+1}}
\chi_6(q)\right]
\right\},
\label{eq:fullbsetype2e}
\end{eqnarray}
\begin{eqnarray}
\chi_6(p)&=&\frac{1}{2}\int \frac{d^3q}{(2\pi)^3} \frac{V(k)}{\omega^2(q)-\frac{\mu^2}{4}}P_{J+1}(\hat{p}\cdot\hat{q})
\nonumber \\&\times&
\left\{\frac{M(q)}{2\bar{\omega}(q)}\sqrt{\frac{J+1}{2J+1}}\left[\sqrt{\frac{J}{2J+1}}\chi_2(q)+
\sqrt{\frac{J+1}{2J+1}}\chi_3(q)\right]
\right.\nonumber \\&&
-\frac{1}{2\bar{\omega}(q)}\sqrt{\frac{J}{2J+1}}\left[q\chi_1(q)+M(q)
\left(\sqrt{\frac{J+1}{2J+1}}\chi_2(q)-
\sqrt{\frac{J}{2J+1}}\chi_3(q)\right)\right]
 \nonumber\\&&
+\frac{M(q)\omega(q)}{\bar{\omega}^2(q)}\sqrt{\frac{J+1}{2J+1}}
\left[q\chi_4(q)+M(q)\left(\sqrt{\frac{J}{2J+1}}\chi_5(q)+
\sqrt{\frac{J}{2J+1}}
\chi_6(q)\right)\right]\nonumber\\&&\left.-
\omega(q)\sqrt{\frac{J}{2J+1}}\left[\sqrt{\frac{J+1}{2J+1}}\chi_5(q)-
\sqrt{\frac{J}{2J+1}}
\chi_6(q)\right]
\right\},
\label{eq:fullbsetype2f}
\end{eqnarray}
\end{subequations}
Obviously in the integrands there appear only four
different linear combinations of the six $\chi_i(p)$
given by the terms in the square brackets.
Thus one can reduce (\ref{eq:fullbsetype2}) to
a coupled system of four
integral equations
\begin{subequations}
\label{eq:bsetype2}
\begin{eqnarray}
\displaystyle
\omega(p)h_1(p)&=&\displaystyle\frac{1}{2}\int \frac{d^3q}{(2\pi)^3} V(k)
\left\{\left(\frac{J}{2J+1}P_{J+1}(\hat{p}\cdot\hat{q})
+\frac{J+1}{2J+1}P_{J-1}(\hat{p}\cdot\hat{q})\right)
\left(h_1(q)+\frac{\mu^2}{4\omega(q)}g_1(q)\right)\right.\nonumber \\&& \displaystyle+
\left.\frac{M(q)}{\bar{\omega}(q)}\frac{\sqrt{J(J+1)}}{2J+1}
\left(P_{J+1}(\hat{p}\cdot\hat{q})-P_{J-1}(\hat{p}\cdot\hat{q})\right)
\left(h_2(q)+\frac{\mu^2}{4\omega(q)}g_2(q)\right)\right\},
\label{eq:bsetype2a}
\end{eqnarray}
\begin{eqnarray}
%
\displaystyle
\left(\omega(p)-\frac{\mu^2}{4\omega(p)}\right)g_1(p)&=&h_1(p)\nonumber 
\\&
+&\displaystyle\frac{1}{2}\int \frac{d^3q}{(2\pi)^3} V(k)\left\{
\frac{pqP_J(\hat{p}\cdot\hat{q})+M(p)M(q)\left(
\frac{J}{2J+1}P_{J+1}(\hat{p}\cdot\hat{q})
+\frac{J+1}{2J+1}P_{J-1}(\hat{p}\cdot\hat{q})\right)}{\bar{\omega}(p)\bar{\omega}(q)}
g_1(q)\right.\nonumber 
\\& +&\displaystyle
\left.\frac{M(p)}{\bar{\omega}(p)}\frac{\sqrt{J(J+1)}}{2J+1}
\left(P_{J+1}(\hat{p}\cdot\hat{q})-P_{J-1}(\hat{p}\cdot\hat{q})\right)g_2(q)\right\}
\label{eq:bsetype2b}
\end{eqnarray}
\begin{eqnarray}
%
\displaystyle
\omega(p)h_2(p)&=&\displaystyle\frac{1}{2}\int \frac{d^3q}{(2\pi)^3} V(k)
\left\{\frac{pqP_J(\hat{p}\cdot\hat{q})+M(p)M(q)\left(
\frac{J+1}{2J+1}P_{J+1}(\hat{p}\cdot\hat{q})
+\frac{J}{2J+1}P_{J-1}(\hat{p}\cdot\hat{q})\right)}{\bar{\omega}(p)\bar{\omega}(q)}\right.
\nonumber \\&& \displaystyle\times
\left(h_2(q)+\frac{\mu^2}{4\omega(q)}g_2(q)\right)\nonumber \\&& \displaystyle+
\left.\frac{M(p)}{\bar{\omega}(p)}\frac{\sqrt{J(J+1)}}{2J+1}
\left(P_{J+1}(\hat{p}\cdot\hat{q})-P_{J-1}(\hat{p}\cdot\hat{q})\right)
\left(h_1(q)+\frac{\mu^2}{4\omega(q)}g_1(q)\right)\right\},
\label{eq:bsetype2c}
%
\end{eqnarray}
\begin{eqnarray}
\displaystyle
\left(\omega(p)-\frac{\mu^2}{4\omega(p)}\right)g_2(p)&=&h_2(p)
+\displaystyle\frac{1}{2}\int \frac{d^3q}{(2\pi)^3} V(k)\left\{\left(
\frac{J+1}{2J+1}P_{J+1}(\hat{p}\cdot\hat{q})+\frac{J}{2J+1}P_{J-1}(\hat{p}\cdot\hat{q})\right)
g_2(q)\right.\nonumber 
\\ &&\displaystyle+
\left.\frac{M(q)}{\bar{\omega}(q)}\frac{\sqrt{J(J+1)}}{2J+1}
\left(P_{J+1}(\hat{p}\cdot\hat{q})-P_{J-1}(\hat{p}\cdot\hat{q})\right)g_1(q)\right\}
\label{eq:bsetype2d}
\end{eqnarray}
\end{subequations}
for the four functions
\begin{equation}
h_1(p)=\frac{\sqrt{\frac{J}{2J+1}}\chi_2(p)+\sqrt{\frac{J+1}{2J+1}}\chi_3(p)}{\omega(p)},
\end{equation}
\begin{equation}
g_1(p)=\frac{\omega(p)\left[h_1(p)+2\frac{p}{\bar{\omega}(p)}\chi_4(p)
+2\frac{M(p)}{\bar{\omega}(p)}\left(\sqrt{\frac{J}{2J+1}}\chi_5(p)
+\sqrt{\frac{J+1}{2J+1}}\chi_6(p)\right)\right]}{\omega^2(p)-\frac{\mu^2}{4}},
\end{equation}
\begin{equation}
h_2(p)=\frac{\frac{p}{\bar{\omega}(p)}\chi_1(p)
+\frac{M(p)}{\bar{\omega}(p)}\left(\sqrt{\frac{J+1}{2J+1}}\chi_2(p)-
\sqrt{\frac{J}{2J+1}}\chi_3(p)\right)}{\omega(p)},
\end{equation}
\begin{equation}
g_2(p)=\frac{\omega(p)\left[h_2(p)+2\left(\sqrt{\frac{J+1}{2J+1}}\chi_5(p)-\sqrt{\frac{J}{2J+1}}\chi_6(p)\right)\right]}{\omega^2(p)-\frac{\mu^2}{4}},
\end{equation}
\end{widetext}
For $J=0$ the Eqs. (\ref{eq:fullbsetype2c},\ref{eq:fullbsetype2d},\ref{eq:fullbsetype2f}) for
$\chi_3$,$\chi_4$ and $\chi_6$ are absent and all terms in the integrands
containing these functions vanish in the remaining three coupled integral equations.
Consequently there exist only the two functions $h_2(p)$ and $g_2(p)$ which
respect the two coupled integral equations (\ref{eq:bsetype2c},\ref{eq:bsetype2d}).

For mesons of category 3 the BSE becomes a coupled system of four integral equations
\begin{widetext}
\begin{subequations}
\label{eq:fullbsetype3}
\begin{eqnarray}
\chi_1(p)&=&\frac{1}{2}\int \frac{d^3q}{(2\pi)^3} \frac{V(k)}{\omega^2(q)-\frac{\mu^2}{4}}P_J(\hat{p}\cdot\hat{q})
\nonumber \\&\times&
\left\{
\omega(q)\chi_1(q)+
\frac{1}{2\bar{\omega}(q)}\left[
q\left(\sqrt{\frac{J}{2J+1}}\chi_2(q)+
\sqrt{\frac{J+1}{2J+1}}\chi_3(q)\right)+M(q)\chi_4(q)\right]
\right\},
\label{eq:fullbsetype3a}\\
\chi_2(p)&=&\sqrt{\frac{J}{2J+1}}\frac{1}{2}\int \frac{d^3q}{(2\pi)^3} 
\frac{V(k)}{\omega^2(q)-\frac{\mu^2}{4}}P_{J+1}(\hat{p}\cdot\hat{q})
\frac{q}{\bar{\omega}(q)}\nonumber \\&\times&
\left\{
\frac{\mu^2}{2}\chi_1(q)+
\frac{1}{\bar{\omega}(q)}\left[
q\left(\sqrt{\frac{J}{2J+1}}\chi_2(q)+
\sqrt{\frac{J+1}{2J+1}}\chi_3(q)\right)+M(q)\chi_4(q)\right]
\right\},
\label{eq:fullbsetype3b}\\
\chi_3(p)&=&\sqrt{\frac{J+1}{2J+1}}\frac{1}{2}\int \frac{d^3q}{(2\pi)^3} 
\frac{V(k)}{\omega^2(q)-\frac{\mu^2}{4}}P_{J-1}(\hat{p}\cdot\hat{q})
\frac{q}{\bar{\omega}(q)}\nonumber \\&\times&
\left\{
\frac{\mu^2}{2}\chi_1(q)+
\frac{1}{\bar{\omega}(q)}\left[
q\left(\sqrt{\frac{J}{2J+1}}\chi_2(q)+
\sqrt{\frac{J+1}{2J+1}}\chi_3(q)\right)+M(q)\chi_4(q)\right]
\right\},
\label{eq:fullbsetype3c}\\
\chi_4(p)&=&\frac{1}{2}\int \frac{d^3q}{(2\pi)^3} 
\frac{V(k)}{\omega^2(q)-\frac{\mu^2}{4}}P_{J}(\hat{p}\cdot\hat{q})
\frac{M(q)}{\bar{\omega}(q)}\nonumber \\&\times&
\left\{
\frac{\mu^2}{2}\chi_1(q)+
\frac{1}{\bar{\omega}(q)}\left[
q\left(\sqrt{\frac{J}{2J+1}}\chi_2(q)+
\sqrt{\frac{J+1}{2J+1}}\chi_3(q)\right)+M(q)\chi_4(q)\right]
\right\}
\label{eq:fullbsetype3d}
\end{eqnarray}
\end{subequations}
Obviously in the integrands there appear only two linear combinations
$\chi_1(q)$ and the term between the square brackets.
Thus one can reduce (\ref{eq:fullbsetype3}) to
a coupled system of two integral equations
\begin{subequations}
\label{eq:bsetype3}
\begin{eqnarray}
\omega(p)h(p)&=&\frac{1}{2}\int \frac{d^3q}{(2\pi)^3} V(k)
\frac{M(p)M(q)P_J(\hat{p}\cdot\hat{q})+
pq\left(\frac{J}{2J+1}P_{J+1}(\hat{p}\cdot\hat{q})
+\frac{J+1}{2J+1}P_{J-1}(\hat{p}\cdot\hat{q})\right)}{\bar{\omega}(p)\bar{\omega}(q)}
\nonumber \\ &&\displaystyle \times
\left(h(q)+\frac{\mu^2}{4\omega(q)}g(q)\right)
,\label{eq:bsetype3a}
%
\\
\left(\omega(p)-\frac{\mu^2}{4\omega(p)}\right)g(p)&=&h(p)
+\frac{1}{2}\int \frac{d^3q}{(2\pi)^3} V(k)
P_J(\hat{p}\cdot\hat{q})
g(q)
\label{eq:bsetype3b}
\end{eqnarray}
\end{subequations}
\end{widetext}
for the two functions
\begin{equation}
h(p)=\frac{\frac{p}{\bar{\omega}(p)}\left(\sqrt{\frac{J}{2J+1}}\chi_2(p)+
\sqrt{\frac{J+1}{2J+1}}\chi_3(p)\right)+\frac{M(p)}{\bar{\omega}(p)}\chi_4(p)}{\omega(p)}
\end{equation}
and
\begin{equation}
g(p)=\frac{\omega(p)\left[h(p)+2\chi_1(p)\right]}{\omega^2(p)-\frac{\mu^2}{4}}.
\end{equation}
For $J=0$ the ansatz for the vertex function contains only component 2. Inserting it into
the BSE gives equation (\ref{eq:fullbsetype3b}) which becomes $\chi_2(p)=0$ in this case.
Thus a meson with quantum numbers $0^{--}$ does not occur in the model with instantaneous
interaction.

For mesons of category 4 the BSE becomes
\begin{equation}
\chi_{1}(p)=\chi_2(p)=0.
\end{equation}
That means that in the model  with instantaneous interaction no mesons of category 4 occur.
The quantum numbers
$J^{+-}$ for $J=2n$ and $J^{-+}$ for $J=2n+1$, respectively, and also $0^{--}$ of category 3
are called exotic.

\section{Appendix C}
Here we define the rest frame meson wave function
and project it onto the quark-antiquark component propagating forward
and backward in time, see \cite{G7} and references cited therein.
 
 The rest frame wave function is
\begin{eqnarray}
&&\!\!\!\!\!\!\!\psi_{JM}^{PC}(\mu,\vec{p})=\nonumber \\
&&\!\!\!\!\!\!\!i\int \frac{dp_0}{2\pi}
S(p_0+\frac{\mu}{2},\vec{p})\chi_{JM}^{PC}(\mu,\vec{p})S(p_0-\frac{\mu}{2},\vec{p}).
\end{eqnarray}
The quark propagator can be decomposed into
\begin{equation}
S(p_0,\vec{p})=S_+(p_0,\vec{p})+S_-(p_0,\vec{p})
\end{equation}
where
\begin{equation}
S_\pm(p_0,\vec{p})=i\frac{T_pP_\pm T_p^\dag\gamma_0}{p_0\mp\omega(p)\pm i\epsilon}
\end{equation}
with  
\begin{equation}
T_p=\exp\left[-\frac{1}{2}\vec\gamma\cdot\hat{p}\left(\frac{\pi}{2}-\varphi_p\right)\right]
\end{equation}
and the projectors
\begin{equation}
P_\pm=\frac{1\pm\gamma_0}{2}.
\end{equation}
Integration over $p_0$
and using $P_\pm\gamma_0=\pm P_\pm$
and $T_p^\dag\gamma_0=\gamma_0T_p$
yields
\begin{eqnarray}
\psi_{JM}^{PC}(\mu,\vec{p})&=&\frac{T_pP_+T_p\chi_{JM}^{PC}(\mu,\vec{p})T_pP_-T_p}{2\omega(p)-M}
\nonumber\\&+&
\frac{T_pP_-T_p\chi_{JM}^{PC}(\mu,\vec{p})T_pP_+T_p}{2\omega(p)+M}.
\label{eq:wfdecomp}
\end{eqnarray}
Due to $T_p^\dag=T_p^{-1}$
the Foldy transformed wave function
$\tilde{\psi}_{JM}^{PC}(\mu,\vec{p})=T_p^\dag\psi_{JM}^{PC}(\mu,\vec{p})T_p^\dag$
has the form 
\begin{equation}
\tilde{\psi}_{JM}^{PC}(\mu,\vec{p})=
P_+\psi_{+JM}^{PC}(\mu,\vec{p})P_-+P_-\psi_{-JM}^{PC}(\mu,\vec{p})P_+,
\end{equation}
with components
$\psi_{\pm JM}^{PC}(\mu,\vec{p})$ propagating 
forward and backward in time. Notice that
for the quark $P_+$ and $P_-$ project out the forward and backward components, respectively,
but for the antiquark it is in the other way around.
The general form of $\psi_{\pm JM}^{PC}(\mu,\vec{p})$ for a meson of given quantum numbers
can be derived by introducing the respective vertex function into (\ref{eq:wfdecomp}).
This yields
\begin{widetext}
\begin{equation}
\psi_{\pm JM}^{PC}(\mu,\vec{p})=\psi_{\pm}(p)Y_{JM}(\hat{p})\gamma_5=
\frac{1}{2}\left[h(p)\pm\frac{\mu}{2}\left(1\pm\frac{\mu}{2\omega(p)}\right)g(p)\right]
Y_{JM}(\hat{p})\gamma_5
\label{eq:wfpm1}
\end{equation}
for mesons of category one,
\begin{eqnarray}
\psi_{\pm JM}^{PC}(\mu,\vec{p})&=&
\left[
\pm\sqrt{\frac{J}{2J+1}}
\psi_{1\pm}(p)+
\sqrt{\frac{J+1}{2J+1}}
\psi_{2\pm}(p)
\right]
\left\{Y_{J+1}(\hat{p})\otimes\vec{\gamma}\right\}_{JM}\nonumber\\ 
&+&
\left[\pm\sqrt{\frac{J+1}{2J+1}}\psi_{1\pm}(p)
-
\sqrt{\frac{J}{2J+1}}\psi_{2\pm}(p)
\right]
\left\{Y_{J-1}(\hat{p})\otimes\vec{\gamma}\right\}_{JM} 
\label{eq:wfpm2}
\end{eqnarray}
with
\begin{equation}
\psi_{i\pm}(p)=
\frac{1}{2}\left[h_i(p)\pm\frac{\mu}{2}\left(1\pm\frac{\mu}{2\omega(p)}\right)g_i(p)\right]
\end{equation}
for mesons of category 2 and
\begin{equation}
\psi_{\pm JM}^{PC}(\mu,\vec{p})=
\mp\psi_{\pm}(p)\left\{Y_{J}(\hat{p})\otimes\vec{\gamma}\right\}_{JM}
=\mp
\frac{1}{2}\left[h(p)\pm\frac{\mu}{2}\left(1\pm\frac{\mu}{2\omega(p)}\right)g(p)\right]
\left\{Y_{J}(\hat{p})\otimes\vec{\gamma}\right\}_{JM}
\label{eq:wfpm3}
\end{equation}
\end{widetext}
for mesons of category 3.

\section{Appendix D}
In this Appendix we show how we numerically solve the BSE.
We calculate the spectra and wave functions with small but
finite values of $\mu_{\rm IR}$, specifically for 
$\mu_{\rm IR}=0.01\sqrt{\sigma}$, $0.005\sqrt{\sigma}$ and $0.001\sqrt{\sigma}$.
As input we use the respective
results for $M(p)$ and $\omega(p)$ which are obtained by an iterative solution
of the gap equation.
With the given small but finite $\mu_{\rm IR}$ the deviation
of $\mu^2(\mu_{\rm IR})$ from the IR limit $\mu^2=\mu^2(0)$
%
turns out to be almost linear
for small values of $\mu_{\rm IR}$.
Then meson mass $\mu$ in the IR limit can be determined
by extrapolation to $\mu_{\rm IR}=0$ with
high confidence. 
\begin{figure}
\includegraphics[width=\hsize,clip=]{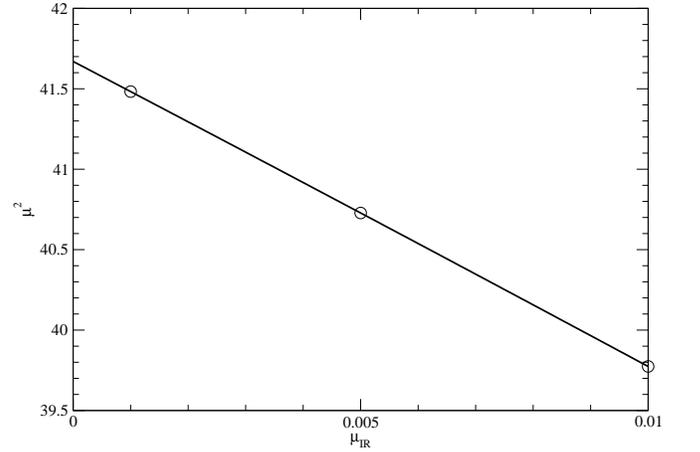}
\caption{Results for the square of the mass of the
fourth radial excitation of the pseudoscalar meson in units of $\sigma$
for three values $0.01\sqrt{\sigma}$, $0.005\sqrt{\sigma}$ and $0.001\sqrt{\sigma}$ 
of the IR regulator $\mu_{\rm IR}$ and extrapolation to the IR
limit $\mu_{\rm IR}=0$.}
\label{fig:pi_0_4_ir}
\end{figure}  
\begin{figure}
\includegraphics[width=\hsize,clip=]{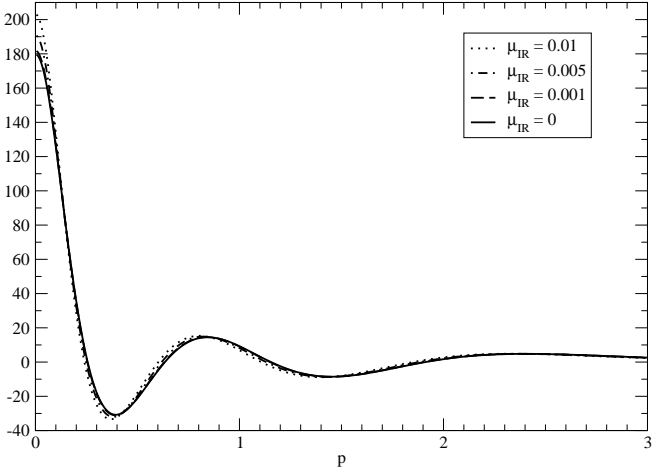}
\caption{Component $\psi_{+}(p)$ of the wave function of the
fourth radial excitation of the pseudoscalar meson
for three small values of $\mu_{\rm IR}$ and
its extrapolation to $\mu_{\rm IR}=0$.
All quantities are given in appropriate units of $\sqrt{\sigma}$.}
\label{fig:pi_0_4_wf}
\end{figure}  
In a similar manner the IR limit of the wave functions can be obtained.
In Figs. \ref{fig:pi_0_4_ir} we show as example 
the determination of the IR limit for the mass of the
fourth radial excitation of the pseudoscalar meson.
The forward-propagating component of the 
wave function of the same state 
is shown in Fig. \ref{fig:pi_0_4_wf}
together with the extrapolated IR limit, which in this plot is hardly
distinguishable from the result with $\mu_{\rm IR}=0.001\sqrt{\sigma}$.

The systems of integral equations
(\ref{eq:bsetype1}), (\ref{eq:bsetype2}) and (\ref{eq:bsetype3})
are numerically solved by expanding the
functions $h,g$ (respectively $h_1,g_1,h_1,g_2$)
in a finite set
$(\varphi_i,i=1,\ldots,N)$ of basis functions.
In the following we show details 
for mesons of category 1: 
With
\begin{subequations}
\label{eq:wfexpansion}
\begin{eqnarray}
h(p)&=&\sum\limits_{i=1}^N c^h_i\varphi_i(p),\\
\frac{\mu}{2} g(p)&=&\sum\limits_{i=1}^N c^g_i\varphi_i(p),
\end{eqnarray}
\end{subequations}
one ends up with
\begin{equation}
A(\mu)\left(\begin{array}{c} \displaystyle \vec{c}^h \\ \displaystyle \vec{c}^g
	\end{array}\right)=
B(\mu)\left(\begin{array}{c} \displaystyle \vec{c}^h \\ \displaystyle \vec{c}^g	\end{array}\right),		
\label{eq:geneigen}
\end{equation}
where the real symmetric $2N\times 2N$ matrices $A(\mu)$ and $B(\mu)$
have the block structure
\begin{eqnarray}
A(\mu)&=&\left(\begin{array}{cc} \displaystyle A^{hh} & \displaystyle 0 \\
			\displaystyle 0 & \displaystyle A^{gg}(\mu) \end{array}\right)
\\
B(\mu)&=&\left(\begin{array}{cc} \displaystyle B^{hh} & \displaystyle B^{hg}(\mu) \\
			\displaystyle B^{gh}(\mu) & \displaystyle B^{gg}(\mu) \end{array}\right)
\end{eqnarray}
with symmetric real $N\times N$ matrices $A^{hh}$, $A^{gg}(\mu)$, $B^{hh}$ and $B^{gg}(\mu)$ and 
real $N\times N$ matrices $B^{hg}(\mu)$ and $B^{gh}(\mu)$ with $B^{hg}(\mu)^T=B^{gh}(\mu)$.
The matrix elements are given by
\begin{equation}
A^{hh}_{ij}=\int \frac{d^3p}{(2\pi)^3} \varphi_i(p)\omega(p)\varphi_j(p),
\label{eq:ahh}
\end{equation}
\begin{equation}
A^{gg}_{ij}(\mu)=\int \frac{d^3p}{(2\pi)^3} \varphi_i(p)
\left[\omega(p)-\frac{\mu^2}{4\omega(p)}\right]
\varphi_j(p),
\label{eq:agg}
\end{equation}
\begin{equation}
B^{hh}_{ij}=\frac{1}{2}\int \frac{d^3pd^3q}{(2\pi)^6} \varphi_i(p)
V(k)P_J(\hat{p}\cdot\hat{q})
\varphi_j(q),
\label{eq:bhh}
\end{equation}
\begin{widetext}
\begin{equation}
B^{hg}_{ij}(\mu)=B^{gh}_{ji}(\mu)=\frac{\mu}{4}\int \frac{d^3pd^3q}{(2\pi)^6}
\varphi_i(p)\frac{V(k)}{\omega(q)}P_J(\hat{p}\cdot\hat{q})\varphi_j(q),
\label{eq:bhg}
\end{equation}
\begin{eqnarray}
B^{gg}_{ij}(\mu)&=&\frac{1}{2}\int \frac{d^3pd^3q}{(2\pi)^6}
\varphi_i(p)V(k)
\nonumber\\ &\times&
\left[\frac{M(p)M(q)P_J(\hat{p}\cdot\hat{q})+
pq\left(\frac{J+1}{2J+1}P_{J+1}(\hat{p}\cdot\hat{q})
+\frac{J}{2J+1}P_{J-1}(\hat{p}\cdot\hat{q})\right)}{\bar{\omega}(p)\bar{\omega}(q)}
+\frac{\mu^2}{4\omega(p)\omega(q)}P_J(\hat{p}\cdot\hat{q})
\right]
\varphi_j(q).
\label{eq:bgg}
\end{eqnarray}
\end{widetext}
One can drop or add terms of order ${\cal O}(\mu_{\rm IR})$
which leads to different results at finite $\mu_{\rm IR}$ but leaves
the IR limit invariant  (see the discussion on the IR limit of the BSE above).
For instance one can drop the $\mu$-dependent terms in $A^{gg}(\mu)$ and $B^{gg}(\mu)$
or replace (\ref{eq:bhg}) by
\begin{equation}
B^{hg}_{ij}(\mu)=B^{gh}_{ji}(\mu)=\frac{\mu}{2}\int \frac{d^3p}{(2\pi)^3}
\varphi_i(p)\varphi_j(p).
\end{equation}
We have checked that these modifications
indeed give --- up to insignificant numerical errors --- the same extrapolated
IR limits.
All results presented
in this paper
have been calculated with the exact expressions (\ref{eq:ahh})--(\ref{eq:bgg})
for finite $\mu_{\rm IR}$.
By introducing a multiplicative parameter $\lambda(\mu)$ on its right hand side,
Eq. (\ref{eq:geneigen}) becomes a generalized eigenvalue problem
which can be solved with standard linear algebra methods yielding
$2N$ eigenvalues $\lambda_i(\mu)$. The masses
$\mu_i$ ($i=1,\ldots,K$) of the $K$ lowest lying mesons with given spin are then found
by solving $\lambda_i(\mu_i)=1$ for the lowest $K$
eigenvalues. The corresponding eigenvectors give the coefficients
in the expansion (\ref{eq:wfexpansion}) in each case.
Finally the functions $h$ and $g$ have to be normalized according to
(\ref{eq:norm13}).
The spectra and wave functions of mesons of categories 2 and 3 are found
in an analogous manner.
As basis functions we take
\begin{equation}
\varphi_i(p)=p^L\exp(-\alpha_i p^2),
\end{equation}
where $L=J$ for mesons of category 1 and 3 and $L=|J-1|$ for mesons of category 2.
With properly chosen parametrs $\alpha_i$ a number of $N=20$ basis functions
is sufficient for a satisfactory numerical accuracy of masses and wave functions
of all states considered in this work.

\end{document}